\documentclass[preprint]{aastex}

\begin{document}

\shortauthors{Dahlstrom et al.}
\shorttitle{Herschel 36 Anomalous DIBs I. Observations}


\title{Anomalous Diffuse Interstellar Bands in the Spectrum of Herschel 36. I. Observations of Rotationally Excited CH and CH$^+$ Absorption and Strong, Extended Redward Wings on Several DIBs\footnotemark}
\footnotetext{Based in part on data obtained from the ESO Science Archive Facility by user DWELTY}

\author{Julie Dahlstrom\altaffilmark{1}, Donald G. York\altaffilmark{2,3}, Daniel E. Welty\altaffilmark{2}, Takeshi Oka\altaffilmark{2,3,4}, L. M. Hobbs\altaffilmark{5}, Sean Johnson\altaffilmark{2}, Scott D. Friedman\altaffilmark{6}, Zihao Jiang\altaffilmark{2}, Brian L. Rachford\altaffilmark{7}, Reid Sherman\altaffilmark{2}, Theodore P. Snow\altaffilmark{8}, and Paule Sonnentrucker\altaffilmark{6}}

\altaffiltext{1}{Carthage College, 2001 Alford Park Dr., Kenosha, WI 53140; jdahlstrom1@carthage.edu}
\altaffiltext{2}{University of Chicago, Astronomy \& Astrophysics Center, 5640 S. Ellis Ave., Chicago, IL 60637}
\altaffiltext{3}{Enrico Fermi Institute}
\altaffiltext{4}{Dept. of Chemistry, University of Chicago, 5735 S. Ellis Ave., Chicago, IL 60637}
\altaffiltext{5}{University of Chicago, Yerkes Observatory, Williams Bay, WI 53191}
\altaffiltext{6}{Space Telescope Science Institute, 3700 San Martin Dr., Baltimore, MD 21218}
\altaffiltext{7}{Embry-Riddle Aeronautical University, Dept. of Physics, 3700 Willow Creek Rd., Prescott, AZ 86301}
\altaffiltext{8}{University of Colorado, CASA -- Campus Box 389, Boulder, CO 80309}

\begin{abstract}

Anomalously broad diffuse interstellar bands (DIBs) at 5780.5, 5797.1, 6196.0, and 6613.6 \AA\ are found in absorption along the line of sight to Herschel 36, the star illuminating the bright Hourglass region of the \ion{H}{2} region Messier 8.  
Interstellar absorption from excited CH$^+$ in the $J=1$ level and from excited CH in the $J=3/2$ level is also seen.  
To our knowledge, neither those excited molecular lines nor such strongly extended DIBs have previously been seen in absorption from interstellar gas.  
These unusual features appear to arise in a small region near Herschel 36 which contains most of the neutral interstellar material in the sight line.
The CH$^+$ and CH in that region are radiatively excited by strong far-IR radiation from the adjacent infrared source Her 36 SE.  
Similarly, the broadening of the DIBs toward Herschel 36 may be due to radiative pumping of closely spaced high-$J$ rotational levels of relatively small, polar carrier molecules.  
If this picture of excited rotational states for the DIB carriers is correct and applicable to most DIBs, the 2.7 degree cosmic microwave background may set the minimum widths (about 0.35 \AA) of known DIBs, with molecular processes and/or local radiation fields producing the larger widths found for the broader DIBs.
Despite the intense local UV radiation field within the cluster NGC 6530, no previously undetected DIBs stronger than 10 m\AA\ in equivalent width are found in the optical spectrum of Herschel 36, suggesting that neither dissociation nor ionization of the carriers of the known DIBs by this intense field creates new carriers with easily detectable DIB-like features.  
Possibly related profile anomalies for several other DIBs are noted.  

\end{abstract}


\section{INTRODUCTION} 
\label{sec-intro}

The diffuse interstellar bands (DIBs) have been observed in the spectra of reddened stars for over 90 years (Heger 1922; Merrill \& Wilson 1938), but remain unidentified.
The known number of DIBs has increased slowly since 1922 and now exceeds 540, to an equivalent width limit of a few m\AA\ for relatively narrow DIBs.
That number may increase, once a complete survey for broad DIBs (with FWHM $>$ 6 \AA) has been accomplished; once some ambiguities between stellar lines and possible DIBs are resolved (Hobbs et al. 2009); and once additional atlases of reddened stars have been completed [to see if the set of DIBs in two atlases of reddened stars, HD 204827 (Hobbs et al. 2008) and HD 183143 (Hobbs et al. 2009), between them, are representative of the general interstellar medium].  
Furthermore, while most known DIBs have been found between 4100 and 9000 \AA, ongoing infrared and ultraviolet surveys for DIBs (Snow \& McCall 2006; Geballe et al. 2011) may reveal additional DIBs outside that wavelength range.
A number of reviews have summarized various observed properties of the DIBs and proposed explanations for their origin (e.g., Herbig 1995; Snow 1997; Fulara \& Kre{\l}owski 2000; Sarre 2006).  

Recent evidence has strengthened the view that the DIBs are due to gas-phase molecules.  
The central wavelengths of the DIBs, from star to star and from observer to observer, referenced to interstellar lines such as those of neutral potassium, agree to within 5 km s$^{-1}$ when the centroid can be measured to that accuracy (i.e., for relatively narrow DIBs; Hobbs et al. 2008).  
Exceptions to that agreement generally can be ascribed to the presence of widely spaced interstellar components, all containing DIBs (Herbig \& Soderblom 1982; Weselak et al. 2010; D. York et al. 2013, in preparation).  
While the DIBs once were thought to arise from defects in interstellar dust grains [see, e.g., Herbig (1995) for a summary], the polarization does not change across the bands (Cox et al. 2011, and references therein), strongly suggesting that the DIBs are not due to the dust.  
A broadening mechanism known as internal conversion has been considered as making it feasible that the carriers are intermediate sized molecules (Danks \& Lambert 1976; Douglas 1977; Smith et al. 1977).  
Such molecules can have closely spaced rotational levels that would lead to wide, nearly structureless features if those rotational levels are populated (e.g., Snow 2002).
Variations in the profiles of emission features seen near the wavelengths of several of the DIBs (and suggested to be due to the DIB carriers) in the Red Rectangle (e.g., Scarrott et al. 1992; Wehres et al. 2011) may reflect changes in the rotational excitation of the carrier molecules with distance from the central star.

Large molecules appear to be widespread in the interstellar medium of the Galaxy.  
Molecules up to mass 64 are known in diffuse molecular clouds (Snow \& McCall 2006); larger molecules, up to mass 185, have been found in dense molecular clouds via radio and submillimeter observations (Kwok 2007); the fullerene C$_{60}$ has been observed in the infrared spectra of a young planetary nebula (Cami et al. 2010) and several reflection nebulae (Sellgren et al. 2010).
Large molecules [e.g., polycyclic aromatic hydrocarbons (PAHs)] are suspected to be the origin of the unidentified infrared bands seen in emission from both Galactic high-latitude cirrus clouds (e.g., Verter et al. 2000; Ricca et al. 2012; c.f. Kwok \& Zhang 2011) and discrete sources.
Such molecules may also be the source of the extended red emission (known especially for the Red Rectangle; Witt \& Boroson 1990) and of the blue fluorescence associated with the central regions of the Red Rectangle (Vijh et al. 2004).   
Moreover, submillimeter spectra of dark clouds have revealed a plethora of unidentified emission lines, many of which may be due to large molecules (e.g., Herbst \& van Dishoeck 2009, and references therein).  
Despite the growing indications that there are a number of large molecular species in the ISM, however, it is uncertain how such large molecules might form, especially as most of the optical and IR features are observed outside of the denser molecular clouds.  
It is thus of interest to pursue the 90-year-old problem of the DIBs -- in hopes of finding specific molecules and broadening mechanisms that might shed light on the chemistry of the interstellar medium.

A number of researchers have tried to find resonance lines or bands of known molecules in the laboratory that correspond to the observed wavelengths of multiple diffuse bands (e.g., Tulej et al. 1998; Motylewski et al. 2000; G\"{u}the et al. 2001; Salama et al. 2011), but no matches have been found for DIBs that are ubiquitous.  
There have been several claims to the detection of ionized naphthalene (Iglesias-Groth et al. 2008, 2012), but if that molecule is present, it is evidently not very widespread (Galazutdinov et al. 2010; Searles et al. 2011; Iglesias-Groth et al. 2012).
Comparisons of laboratory spectra of {\it l}-C$_3$H$_2$ with the current list of DIBs yielded intriguing possible matches with the broad DIBs at 4881.1 and 5450.6 \AA\ (Maier et al. 2011), but the agreement is not as good for several of the other measured bands and the inferred column density of {\it l}-C$_3$H$_2$ is much higher than the values obtained from radio observations of that molecule (Kre{\l}owski et al. 2011; Liszt et al. 2012).
 
Although it is well known that the profiles of different DIBs are characterized by different shapes and widths, the cause(s) of those differences generally are not understood.  
While some DIB profiles appear to be symmetric, others are decidedly asymmetric; the FWHM of the profiles range from $\sim$0.35 \AA\ to more than 40 \AA.  
Moreover, spectra obtained over the past 20 years have hinted at variations in the profiles of individual DIBs in different sight lines. 
Because the profile variations are in many cases somewhat subtle, and because the DIBs can be broad and/or weak, both relatively high spectral resolution and high signal-to-noise (S/N) ratios are generally required for their detection.  
In some cases, the observed profiles reflect the complex structure of the ISM -- as seen in the narrower atomic and molecular absorption lines along most sight lines (e.g., Herbig \& Soderblom 1982; Weselak et al. 2010) -- but that does not account for all such differences.  
To the extent to which variations in the DIB profiles are due to specific environmental differences, they may yield useful diagnostics of the physical conditions where the carriers reside.  
Determination of the intrinsic shapes and widths of the DIB profiles would aid in narrowing the search for specific carriers responsible for the DIBs.

High-resolution, high-S/N ratio spectra have revealed substructure (and variations in that substructure) in a number of the DIBs (Sarre et al. 1995; Kre{\l}owski \& Schmidt 1997; Kerr et al. 1998; Galazutdinov et al. 2002, 2008; Slyk et al. 2006), though on velocity scales much smaller than those discussed later in this paper.  
Ehrenfreund \& Foing (1996) interpreted the structure seen in the commonly observed portions of the $\lambda\lambda$5797.1, 6379.2, and 6613.6 DIBs as representing rotational contours of large gas-phase molecules.  
In this picture, the subcomponents of each DIB should both broaden and increase in separation with increasing rotational temperature.  
Webster (1996) proposed instead that the substructure could be due to substitution of $^{13}$C atoms for $^{12}$C in large carbonaceous molecules, with an average of 1.2 such substitutions (and an overall Poisson distribution of substitutions) giving a reasonable fit to the $\lambda$6613.6 DIB. 
In that case, the subcomponent separations (of order 0.3 \AA\ for the $\lambda$6613.6 DIB) should be invariant, but the subcomponent widths should increase as the square root of the kinetic temperature.  
Webster (2004) also suggested an analogous zero-point vibrational isotope shift as a means of broadening some of the narrower DIBs (e.g., the $\lambda$6196.0 DIB).  
Fits to the profiles of the $\lambda\lambda$5797.1, 6196.0, 6379.2, and 6613.6 DIBs by Walker et al. (2000, 2001) seemed generally consistent with isotopic substitution, but a slightly larger separation between subcomponents was found for HD 37061, and an additional ``anomalous'' broadening was required for the $\lambda$6379.2 DIB.  
Subsequent modeling of the $\lambda$6613.6 DIB by Cami et al. (2004), however, found slight relative shifts of the main red and blue subcomponents in different sight lines -- more consistent with rotational structure. 

Here, we report the discovery of highly broadened DIBs in the spectrum of the star Herschel 36, with pronounced wings extended redward from the central cores of the features.  
In this line of sight, absorption is also found from the $J=1$ excited level of CH$^+$ and from the $J=3/2$ excited level of CH.  
To our knowledge, neither those excited molecular lines nor such strongly extended DIBs have previously been seen in absorption from interstellar gas.  
A strong, nearby infrared source (Goto et al. 2006) could explain these phenomena by infrared pumping of rotational levels in the ground states of both the DIB carrier molecules and the CH$^+$ and CH (Oka et al. 2013).  
Investigation of these anomalous DIB profiles -- and of the local environmental conditions in which they arise -- may thus provide both new information on the broadening mechanisms and new constraints on the identities of the DIB carriers.

In the rest of this paper, we first briefly outline in Section~\ref{sec-region} the morphology and characteristics of the central region of the nebula Messier 8, near the Hourglass Nebula, which is centered $\sim$15 arcsec from Herschel 36.  
The observations of DIBs, CH$^+$, and CH toward Herschel 36 and several other stars in that region are described in Section~\ref{sec-obs}.  
In Section~\ref{sec-disc}, we discuss the size and properties of the local region near Herschel 36 where those anomalous interstellar features arise and we relate those features to observations of DIB profile variations in other sight lines.
Section~\ref{sec-conc} summarizes our conclusions.  
Detailed analysis of the CH$^+$ excitation and simple molecular models that can match the shapes of the extended redward wings seen for several of the DIBs are discussed in a companion paper (Oka et al. 2013; hereafter Paper II).  
Those models suggest that the carriers of the DIBs exhibiting strong redward wings must be relatively small polar molecules.
Throughout this paper, we use DIB rest wavelengths determined from the spectrum of HD 204827 (Hobbs et al. 2008; see Friedman et al. 2011):  the wavelengths are in air and are rounded to one decimal place.

\section{THE NEBULA MESSIER 8 AND THE STAR HERSCHEL 36} 
\label{sec-region}

\begin{figure}[b!] 
\plottwo{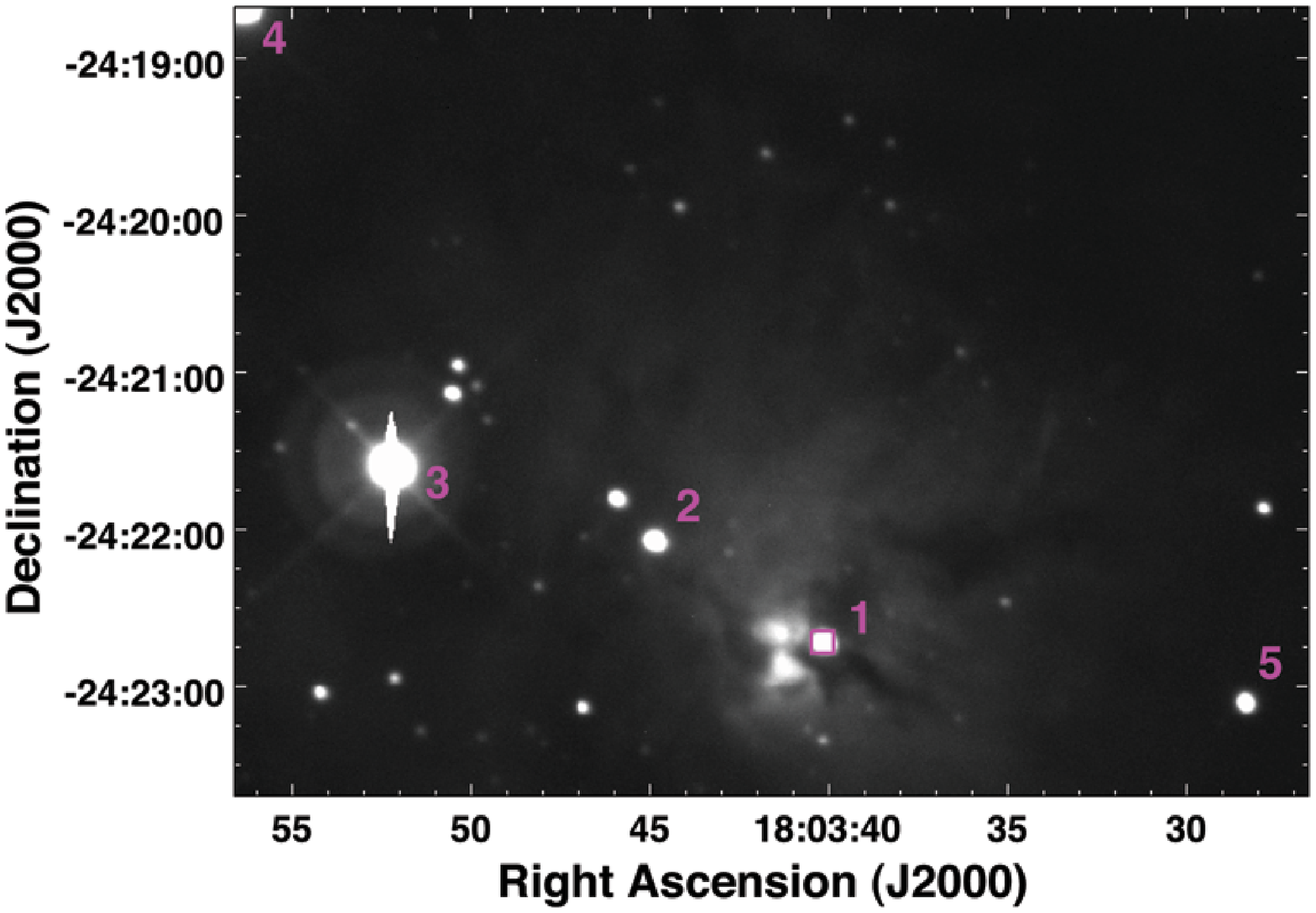}{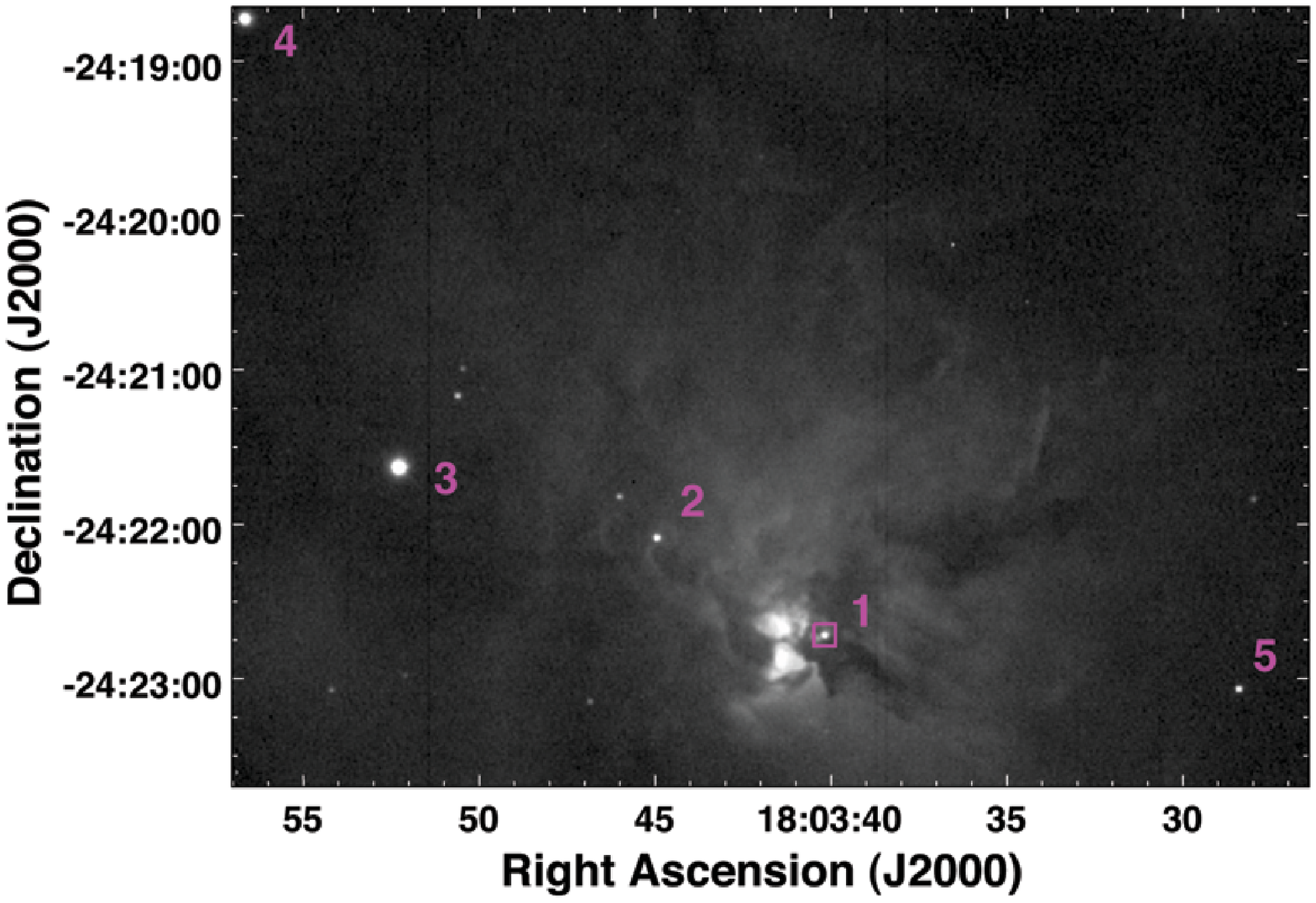}
\caption{The region around Herschel 36, imaged with no filter ({\it left}) and through an H$\alpha$ filter ({\it right}).  
North is up; East is to the left.
Star~1 = Herschel~36 (O7.5~V); star~2 = CD$-$24~13810 (B2~V); star~3 = HD~164794 (9~Sgr; O4~V); star~4 = HD~164816 (O9.5~III-IV(n)); star~5 = CDZ115 (K5~III); all appear to be members of the young cluster NGC 6530.
The small box around Herschel~36, to the right of the bright Hourglass Nebula, is the 8.5 arcsec field of the infrared composite image given in Figure~\ref{fig:goto}.
A dark lane, perhaps 2--3 arcsec ($\sim$4000 AU) wide and shadowed in both continuum and H$\alpha$ emission, appears to cross Herschel~36 from the WSW. 
The H$\alpha$ exposure gives an unsaturated view of the distribution of H$\alpha$ emission in the region and shows the dominant effect of Herschel~36 on the Hourglass H$\alpha$ emission.   
The left-hand (unfiltered) image is a portion of a 20-sec exposure from the PROMPT3 telescope (Reichart et al. 2005); the scale is 0.59 arcsec per pixel.
The right-hand (H$\alpha$) image is from the PROMPT2 telescope; the scale is 0.41 arcsec per pixel.
Both images were requested remotely with the assistance of Vivian Hoette, Yerkes Observatory.}
\label{fig:prompt}
\end{figure}

The star Herschel 36 [RA $18^h03^m40\fs3$ ; DEC $-24\degr22\arcmin43\arcsec$ (J2000)] was observed as part of our extensive, high S/N ratio survey of spectra of hot stars, in search of correlations of DIBs with other interstellar parameters (Thorburn et al. 2003; Friedman et al. 2011).  
A triple system probably consisting of three O stars (Arias et al. 2010), Herschel 36 (Herschel 1847), is a member of the very young star cluster NGC 6530. 
Two other cluster O stars of still earlier spectral type, HD 164794 (9 Sgr) and HD 165052, appear to be primarily responsible for exciting the large, resulting \ion{H}{2} region M8 (NGC 6523; the Lagoon Nebula; Tothill et al. 2008).  
Owing to its proximity, however, Herschel 36 is primarily responsible for ionizing the brightest part of M8, the Hourglass Nebula (Thackeray 1950), which appears to be a ``blister'' on the central molecular cloud (Woodward et al. 1986).  
The key stars and some associated nebular structures are shown in Figure~\ref{fig:prompt}.  
Estimates of the distance to the cluster range from 1.25 to 1.8 kpc (e.g. van den Ancker et al. 1997; Arias et al. 2006); we adopt 1.5 kpc.  
Most cluster members, including the exciting stars HD 164794 (9 Sgr) and HD 165052, show a fairly uniform, foreground reddening near $E(B-V)$ = 0.33 (with some variations; Walker 1957), fairly typical far-UV extinction curves, and an average value for the total-to-selective extinction $R_{\rm v}$ = $A_{\rm v}$/$E(B-V)$ of about 3.7 (Jenniskens \& Greenberg 1993; Patriarchi \& Perinotto 1999; Fitzpatrick \& Massa 2007).  
In contrast, Herschel 36 shows additional, evidently very local reddening that boosts its color excess to $E(B-V)$ = 0.87, and exhibits a very flat far-UV extinction curve, with a weaker than average 2175 \AA\ bump (Fitzpatrick \& Massa 1990, 2007, 2009).  
The peculiar extinction curve of Herschel 36 leads to estimates of $R_{\rm v}$ that range from 5.2 to 6.0 (Hecht et al. 1982; Arias et al. 2006; Fitzpatrick \& Massa 2007, 2009), among the highest values known.  
The additional, local absorption toward Herschel 36 thus amounts to almost $A_{\rm v}$ = 4 mag, compared to the values found for other cluster stars (including one at a projected distance of less than 1 pc). 

\begin{figure}[b!]
\epsscale{0.8}
\plotone{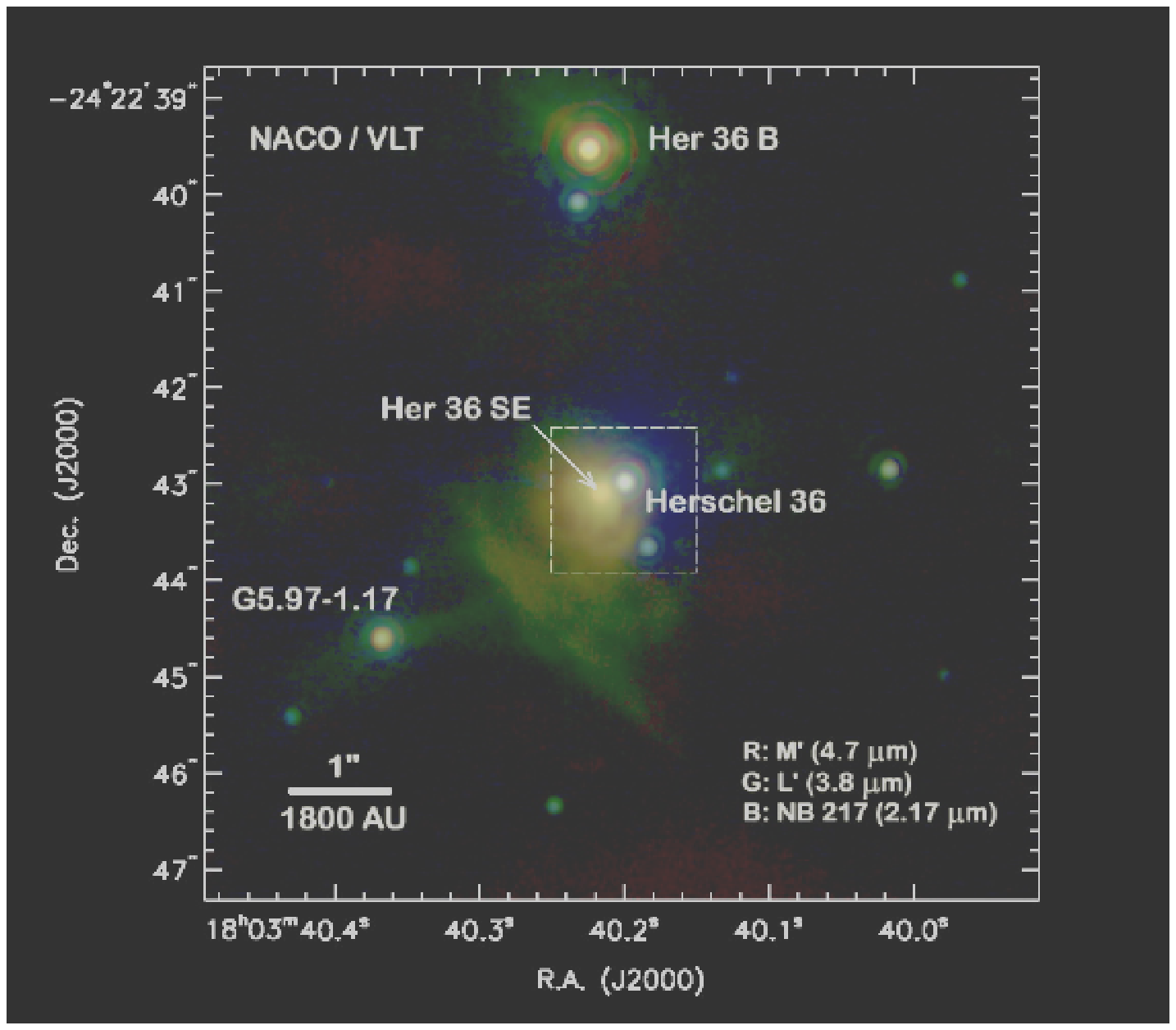}
\caption{Multiband near-IR image of the region very close to Herschel 36, reproduced from Goto et al. (2006).
North is up; East is to the left.
The color coding is explained in the legend; the contrast has been slightly adjusted to bring out the faint reddish emission.
The field includes two bright infrared sources:  Her 36 SE, 0.25 arcsec (375 AU) southeast of Herschel 36, and Her 36B, 3.6 arcsec (5400 AU) north of Herschel 36 (assuming a distance of 1.5 kpc).  
The greenish central nebulosity may correspond to a portion of the dust lane apparent in the optical images shown in Figure~\ref{fig:prompt}.
(reproduced with permission of the AAS)}
\label{fig:goto}
\end{figure}

The immediate vicinity of Herschel 36 and the Hourglass exhibits intricate structure, reflecting a complex distribution of gas and dust acted on by the intense radiation from Herschel 36 (see, e.g., the {\it HST}/WFPC2 image shown in Fig. 9 of Tothill et al. 2008).
Very strong emission from the $J=2-1$, $J=3-2$, and $J=4-3$ rotational transitions of CO has been observed within 1 arcmin of Herschel 36 (White et al. 1997).
The relatively weak CO absorption seen in the {\it FUSE} spectrum of Herschel 36 suggests, however, that much of the CO emission may be from material located behind Herschel 36 (B. Rachford et al., in preparation).
Emission from the 1-0 S(1) transition of H$_2$ at 2.12 $\mu$m has also been detected, with velocities ranging over about 70 km s$^{-1}$ in an apparent bipolar structure roughly centered on Herschel 36 and oriented SE to NW (Burton 2002).
The morphology and velocity range exhibited by the excited H$_2$ emission suggest that the excitation may be due to shocks driven by a bipolar outflow from Herschel 36, though fluorescence (in the strong local UV radiation field) may also contribute (Burton 2002).
A supposed ultracompact \ion{H}{2} region (G5.97-1.17) found within 3 arcsec of Herschel 36 may instead be a nearby circumstellar disk being photoevaporated by Herschel 36 -- perhaps a very distant proplyd (Stecklum et al. 1998).

A multiband near-IR image of the immediate vicinity of Herschel 36 is shown in Figure~\ref{fig:goto} (reproduced from Goto et al. 2006).
The IR source Her 36 SE, highlighted by Goto et al. (2006), lies just 0.25 arcsec SE of Herschel 36.  
The underlying source for Her 36 SE is completely obscured, but is inferred to be an early B star with $A_{\rm v}$ $>$ 60 mag -- deeply embedded in dense, warm dust and powering a very compact \ion{H}{2} region.  
The more extended dust is perhaps being violently removed by the influence of Herschel 36 (Goto et al. 2006).  
A cluster of likely pre-main-sequence stars of low to intermediate mass, most of which have not been detected at visible wavelengths, is found in K-band images of the Hourglass Nebula region (Arias et al. 2006).  
The presence of this cluster reinforces the evident similarity between the central regions of M8 and of M42, the Orion nebula (Goto et al. 2006), although M8 is more distant from the Sun by a factor of at least three.  
M42 is similarly associated with a blister on the background molecular cloud, has several hot stars very nearby and has a cluster of primarily low-mass stars associated with it. 
Also shown in Figure~\ref{fig:goto} is a nebulosity, bright at 3.8 $\mu$m, which is centered 0.7 arcsec SE of Herschel 36 (and adjacent to the star), about 1.4 arcsec in extent to the SW and running about 3 arcsec in the SW to NE direction.  
That nebulosity may correspond to an illuminated section of the long dust lane, visible in the optical images in Figure~\ref{fig:prompt}, that runs WSW to ENE.
If at the distance of Herschel 36, that feature has dimensions of about 2100 by 6600 AU; it may have a temperature of a few hundred Kelvin.  
As shown in Figure~\ref{fig:prompt}, the Hourglass is to the left of this 3.8 $\mu$m nebulosity.  
The rest of the dust lane, seen in shadow in Figure~\ref{fig:prompt}, must be cold, as it does not appear in Figure~\ref{fig:goto}.  
Intermediate temperature nebulosities may also be seen (in red) in the vicinity.  

\section{DIBS AND MOLECULAR LINES IN THE SPECTRUM OF HERSCHEL 36}
\label{sec-obs}

\subsection{Optical Spectra of Herschel 36}
\label{sec-obsH36}

Optical spectra of Herschel 36 were obtained with the Apache Point Observatory (APO) ARC echelle spectrograph (ARCES; Wang et al. 2003), which covers the wavelength range from about 3800 to 11000 \AA\ at a resolution of about 8 km s$^{-1}$.  
The observations were restricted to within two hours of the meridian crossing of the star, because of the low declination and consequent high airmass for this star when observed from APO.  
Thirty-five spectra, each 1/2 hour long and reaching S/N $\sim$ 100, were taken over three nights in the summer of 2001, thirteen nights in the summer of 2002, and one night in the summer of 2005.  
The individual spectra were processed as described in Thorburn et al. (2003), then combined.
The co-added spectrum has a signal-to-noise ratio of about 600 per 0.17 \AA\ resolution element near 6500 \AA, somewhat less than the typical S/N $\sim$ 1000 achieved for stars in our program.  
The velocity zero point is set by the strongest component seen in the interstellar \ion{K}{1} line ($\lambda_{\rm rest}$ = 7698.965 \AA).
The differential velocity errors across an entire ARCES spectrum are typically less than 1 km s$^{-1}$, and the absolute error in the velocity scale is dominated by any unresolved component structure in the \ion{K}{1} line in the line of sight.

\begin{figure}[b!]
\plotone{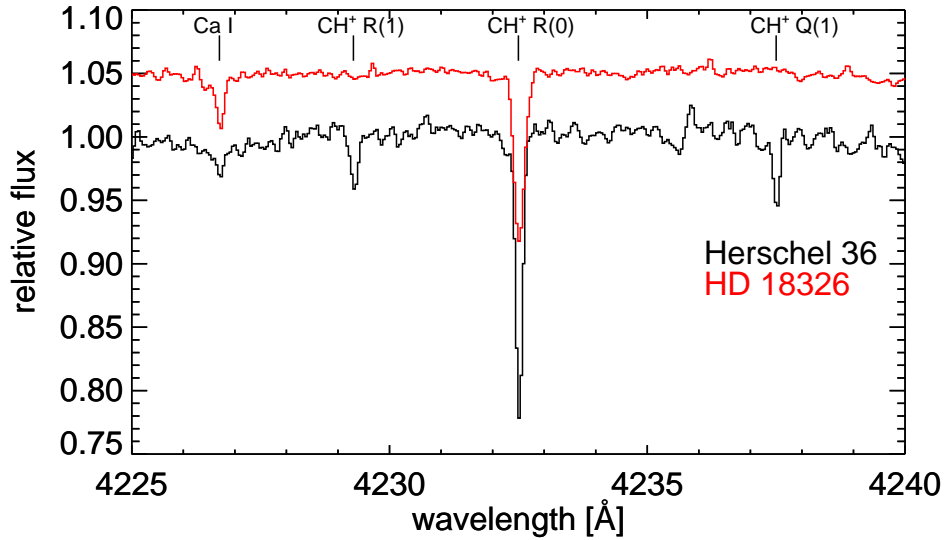}
\caption{Comparison of the CH$^+$ A-X (0,0) band spectra toward Herschel 36 and HD 18326, observed with the ARC echelle spectrograph at the Apache Point Observatory.  
The observed R(1) and Q(1) lines are indicated by tick marks; this is the first detection of these rotationally excited lines in the ISM.  
The two stars have the same spectral type and roughly similar values of $E(B-V)$, but the visual absorption is much higher for Herschel 36, since it has an $R_{\rm v}$ value near 6.  
The wavelength region shown is evidently free of significant stellar spectral lines and of DIBs.}
\label{fig:chp}
\end{figure} 

Examination of the averaged spectra in the summer of 2011 showed immediately that the DIB at 5780.5 \AA\ was anomalously broad and that absorption from the excited $J=1$ level of CH$^+$ was present for the A-X (0,0) band near 4232 \AA\ (Fig.~\ref{fig:chp}).  
Inspection of other DIBs revealed that those at 5797.1, 6196.0, and  6613.6 \AA\ also showed enhanced redward wings, extending to more than twice the normal widths of those DIBs seen in the spectra of other sight lines in our archive.  
The extended redward wings on those DIB profiles were then confirmed by examination of lower S/N spectra of Herschel 36 recorded using the University College London Echelle Spectrograph (UCLES; FWHM $\sim$ 5 km s$^{-1}$) on the Anglo-Australian Telescope (taken by DEW in 1998); of higher S/N spectra from the Fiber-fed Extended Range Optical Spectrograph (FEROS; FWHM $\sim$ 6.25 km s$^{-1}$; Kaufer et al. 1999) at the European Southern Observatory (ESO), recorded from 2006 to 2009 and retrieved from the ESO archive\footnotemark\ in the fall of 2011; and of spectra from the Magellan MIKE spectrograph (FWHM $\sim$ 6 km s$^{-1}$) obtained in 2012 July (by SJ and DEW).
\footnotetext{http://archive.eso.org/wdb/wdb/eso/repro/form}
Weighted sums of 22 FEROS spectra (total exposure $\sim$14.7 hours) and of nine MIKE spectra (total exposure $\sim$1.4 hours) both yielded S/N $\sim$ 1000 per resolution element near 6600 \AA.
The higher S/N FEROS and MIKE spectra also provided more accurate measurements of the rotationally excited molecular lines -- including those from the CH$^+$ A-X (1,0) band near 3957 \AA\ and the CH A-X (0,0) band near 4300 \AA\ (Paper II) -- as well as more stringent limits on CN absorption and some discrimination between stellar and interstellar features.
 
\subsection{Anomalous Diffuse Interstellar Bands in the Spectrum of Herschel 36}
\label{sec-obsdibs}

\begin{figure}[b!] 
\plotone{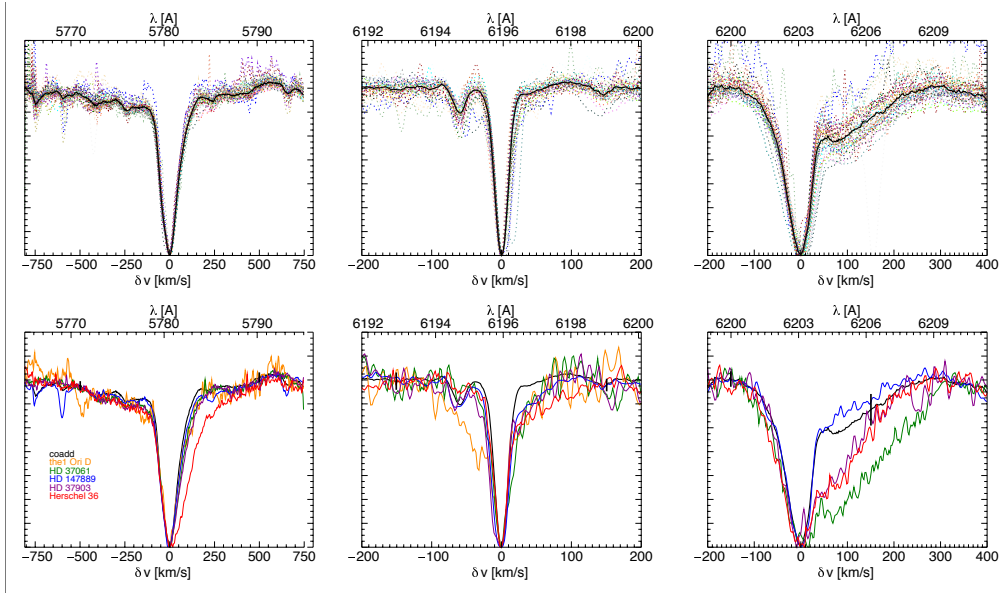}
\caption{ARCES profiles for the DIBs at 5780.5, 6196.0, and 6203.1--6204.8 \AA.
The top panels show the spectra for 47 O and B stars with $E(B-V)$ $>$ 0.35 (faint dashed lines) and the mean spectra (solid black lines; obtained with 3-$\sigma$ clipping of discrepant points). 
In each case, the spectra are scaled to match core depths.  
A number of weaker DIBs may also be discerned in the three mean spectra.  
The dashed traces that deviate in places from the main cluster of points are stellar lines. 
The bottom panels reproduce the mean profile for each DIB (black lines), together with the specific DIB profiles for five stars of interest: Herschel 36 (red), HD 37061 (green), HD 147889 (blue); HD 37903 (violet), and $\theta^1$ Ori D (gold).  
Black vertical bars centered on the mean spectra in the bottom panels show the scatter at those wavelengths among the individual spectra in the top panels. 
For the rightmost bottom panel, $\theta^1$ Ori D is not shown because the normalization is spoiled by stellar lines.}
\label{fig:meandib}
\end{figure}

The uniqueness of the DIB profiles toward Herschel 36 may be seen in Figure~\ref{fig:meandib}, which shows a composite of ARCES spectra of 47 O and B stars for the DIBs at 5780.5, 6196.0, and 6203.1--6204.8 \AA, with the individual spectra shown as dotted lines and the mean spectra (with 3-$\sigma$ outliers removed) as solid lines.  
For each DIB, the profiles for all stars were scaled to a common central depth.
Three stars -- Herschel 36, HD 37903, and HD 37061 -- show extensive wings beyond the typical line widths, for all three DIBs.  
While the wings are strongest toward Herschel 36 for the DIBs at 5780.5 and 6196.0 \AA, the DIB at 6204.8 \AA\ -- which could be either a wing of the DIB at 6203.1 \AA\ or a separate DIB (see discussion below) -- is strongest toward HD 37061.  
More subtle wings may be discerned for some of the other stars shown in the Figure.  

\begin{figure}[b!]
\plotone{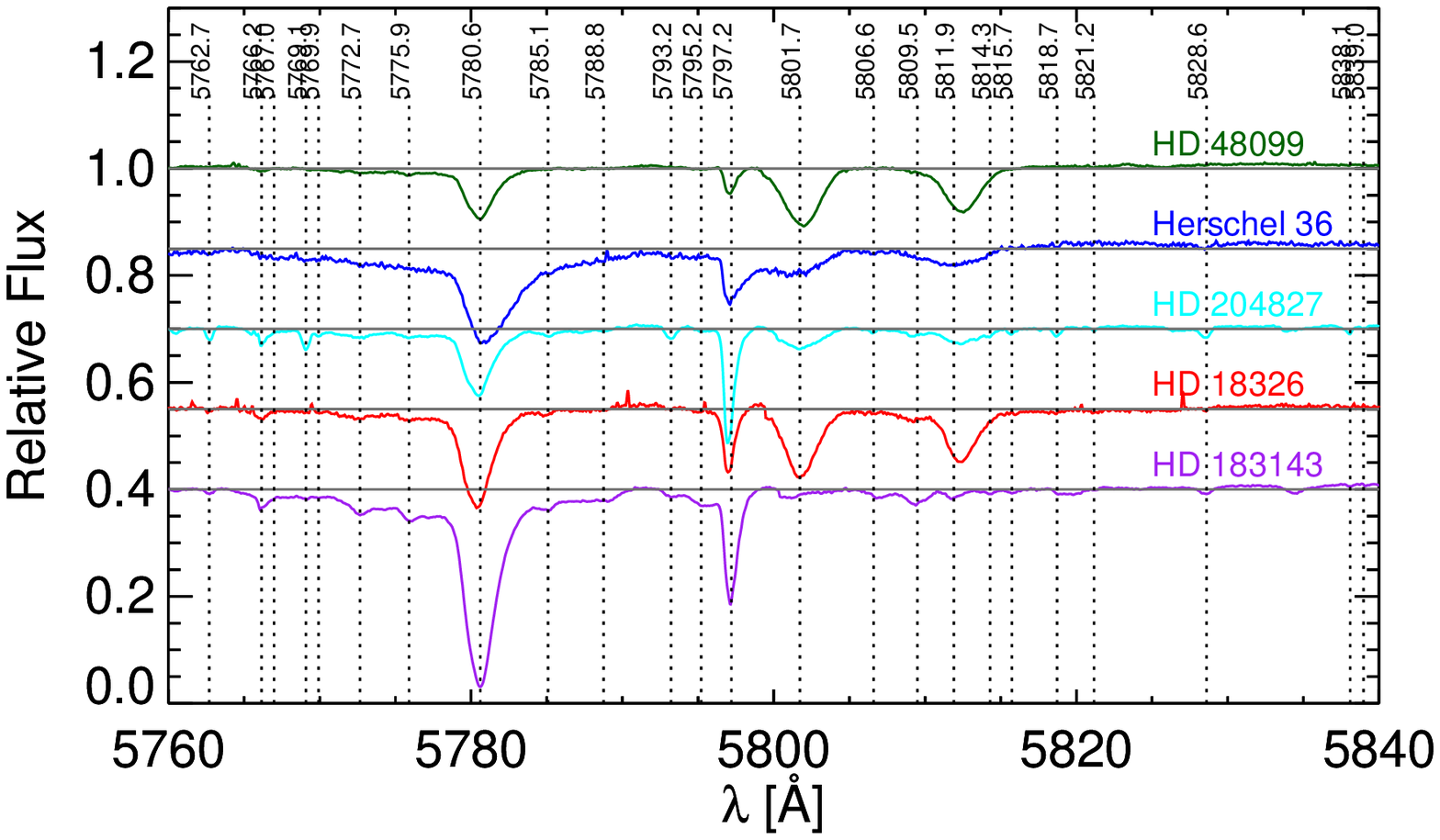}
\caption{ARCES spectra near the $\lambda$5780.5 and $\lambda$5797.1 DIBs, for four stars earlier than B1 and one later type star. 
The five stars cover ranges in reddening and interstellar line strengths; the depths are not rescaled to match at $\lambda$5780.5.  
The dashed vertical lines indicate the positions of known DIBs.
Note the variations in the strengths of both the broad feature from 5765 to 5790 \AA\ and the many narrower DIBs within that broad feature.
The $\lambda$5780.5 and $\lambda$5797.1 DIBs are both anomalously broad toward Herschel 36. 
The two strong lines to the right of the $\lambda$5797.1 DIB for the four hottest stars (but not for the B7 Ia star HD 183143) are stellar features of \ion{C}{4}.}
\label{fig:5stars}
\end{figure} 

In Figure~\ref{fig:5stars}, the normalized ARCES spectra of the spectral region 5740 to 5840 \AA\ are shown for five stars: HD 48099 [O6.5 V(n)((f)), $E(B-V)$ = 0.27], Herschel 36 [O7.5 V, $E(B-V)$ = 0.87], HD 18326 [O6.5 V(n)((f)), $E(B-V)$ = 0.7], HD 183143 [B7 Ia, $E(B-V)$ = 1.27] and HD 204827 [B1 V, $E(B-V)$ = 1.11] -- illustrating both some of the variations in the absorption seen in different sight lines and potential complications from stellar lines for some spectral types.
The stars HD 48099 and HD 18326 have spectral types similar to Herschel 36, while HD 183143 and HD 204827 have higher reddening (and higher signal to noise spectra).  
Absorption from C$_2$, C$_3$, CN, and CO is quite strong toward HD 204827, but very weak toward HD 183143 (e.g., Hobbs et al. 2008, 2009; T. P. Snow, private communication).
The main interstellar cloud toward Herschel 36 is likely subject to very strong IR and UV radiation (\S\S~\ref{sec-region} and \ref{sec-local}).
The two strong absorption features to the right of the $\lambda$5797.1 DIB for the four hottest stars are stellar lines of \ion{C}{4}. 
Comparisons of spectra of Herschel 36 from different epochs indicate clear velocity shifts of the \ion{C}{4} lines (in that triple star system) with respect to the stationary $\lambda$5797.1 DIB, confirming their stellar nature.

It is evident that the entire region is complex, possibly containing a blend of DIBs (e.g., Jenniskens \& D\'{e}sert 1993; Kre{\l}owski \& Sneden 1993).  
The broad feature between 5765 and 5790 \AA\ that underlies the strong $\lambda$5780.5 DIB is most clearly seen for HD 183143 and Herschel 36, but may be absent for the comparably reddened HD 204827.
Since that broad feature is asymmetrically located with respect to the $\lambda$5780.5 DIB, there is no indication that it is related to lifetime broadening of that DIB, but its presence would hinder the discernment of such a feature if it were to exist.  
The narrow (FWHM $\la$ 2 \AA) features superposed across that broad feature, seen clearly in the spectra of HD 183143 and HD 204827, are not as apparent toward Herschel 36.  
The strongly extended wings to the red, for both the $\lambda$5780.5 and $\lambda$5797.1 DIBs, can be seen for Herschel 36, but not for the other stars.
Since the origin of the DIBs is unknown, there is no way to know if these various empirical features are related or if their separate manifestations just overlap in the spectra -- but they should all be distinguished in quantitative studies.  

\begin{figure}[b!]
\plotone{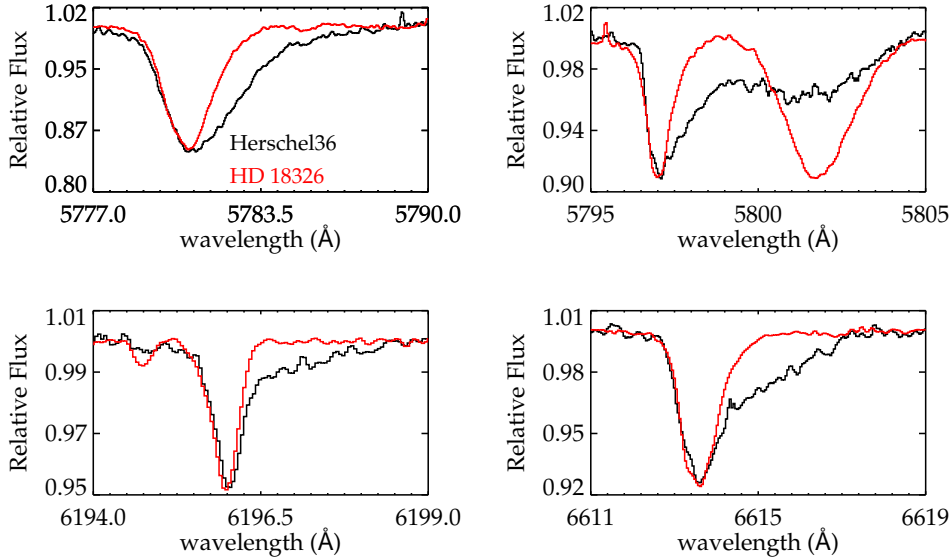}
\caption{Comparison of the $\lambda\lambda$5780.5, 5797.1, 6196.0, and 6613.6 DIBs toward Herschel 36 (black) and HD 18326 (red), observed with ARCES.  
The two stars are well matched in both spectral type and reddening; the spectra have been scaled to match the depths of the DIBs at the line cores.  
All four DIBs show extended redward wings toward Herschel 36, but not toward HD 18326.  
The broad feature to the right of the $\lambda$5797.1 DIB is a stellar \ion{C}{4} line (Figure~\ref{fig:5stars}).}
\label{fig:4dibs}
\end{figure} 

Figure~\ref{fig:4dibs} shows the profiles of the four DIBs on which we concentrate, here over-plotted for HD 18326 and Herschel 36 (of similar spectral type and reddening).  
The profiles of the four DIBs show both similarities and differences.  
Toward HD 18326, a significant intermediate velocity component (seen in \ion{Na}{1} and \ion{Ca}{2}) produces a very weak blueward wing on the profiles of the narrower DIBs, but the cores of the profiles are fairly symmetrical.
Toward Herschel 36, the profiles of all four DIBs rise fairly steeply on the short wavelength side; the DIBs at 5780.5 and 5797.2 \AA\ then rise less steeply to the long wavelength side to form the strong, extended redward wings.  
The $\lambda$6196.0 and $\lambda$6613.6 DIBs, however, have cores that are fairly symmetrical (as for HD 18326), but the profiles then break about half way (or further) up the long wavelength side to make the extended red wings.  
For all four of these DIBs, the redward wings appear to be fairly smooth, with no obvious signs of substructure (as seen in higher resolution spectra of the cores of some DIBs; e.g., Kerr et al. 1998; Cami et al. 2004; Galazutdinov et al. 2008).
Higher resolution spectra of Herschel 36 will be obtained to explore the profiles in more detail.

The additional weak components seen at more negative velocities in \ion{Na}{1} and \ion{Ca}{2} toward Herschel 36 (\S~\ref{sec-neigh}) do not appear to affect the short wavelength sides of DIB profiles there, and there are no components at velocities corresponding to the redward wings seen for the DIBs.  
Likewise, there appears to be no significant absorption at large positive velocities in the far-UV lines of \ion{Fe}{2} or \ion{N}{1} (B. Rachford et al., in preparation), and the positive-velocity emission from excited H$_2$ is found well to the NW of Herschel 36 (Burton 2002).
The redward DIB wings thus appear not to be a product of complex interstellar component structure toward Herschel 36.

The differences in the redward wings in the profiles of the DIBs toward Herschel 36 ($\lambda$5780.5 and $\lambda$5797.1 versus $\lambda$6196.0 and $\lambda$6613.6) are too marked to be caused by issues with the stellar continuum or inadequate S/N (though those factors may contribute).  
Presumably, they must arise from differences in the structure of the molecules that are carriers of the DIBs.  
As pointed out in Paper II, this probably means that those DIBs do not arise from a common carrier.  
The peculiar properties of the region very near a given star (e.g., intensities of UV and IR radiation) may lead to some DIB carriers existing in different relative amounts there than in the general medium [as defined by the mean relationships shown by Friedman et al. (2011); see \S~\ref{sec-disc}].
Such differences in the relative abundances of the DIB carriers in the general foreground and in the small, local region around Herschel 36 may thus lead to differences in the overall shapes of the DIB profiles.  
In Paper II, it is suggested, following an analysis of CH$^+$, that the DIBs exhibiting extended wings to the red toward Herschel 36 are due to relatively small (3--6 heavy atoms) molecules with large dipole moments.  
The requirement that the carrier molecules of those DIBs be polar allows a number of suggested candidate carriers to be ruled out.

\subsection{Profiles and Strengths of Other DIBs toward Herschel 36}
\label{sec-otherdibs}

In light of the unusual profiles seen for several of the well-studied DIBs toward Herschel 36, it would be of interest to look for anomalies in the profiles and/or strengths of other DIBs in that sight line.
In Table~\ref{tab:3los}, we compare the strengths and profiles of some of the DIBs detected in the spectrum of Herschel 36 with those seen toward HD 183143 and HD 204827 (Hobbs et al. 2008, 2009; see also \S~\ref{sec-obsdibs}).   
The table lists 45 DIBs detected toward all three stars, with measured equivalent width $>$ 20 m\AA\ for at least one of the two atlas stars, FWHM $<$ 4 \AA, and (in most cases) no significant blending with other stellar or interstellar features.
In most cases, the DIBs toward Herschel 36 are either weaker than those toward HD 183143 and HD 204827 or else of intermediate strength.
Some of the weaker DIBs (e.g., $\lambda\lambda$5418.9, 6113.2, 6439.5, 6445.3, 6660.7) are much weaker than for the two atlas stars -- even though $E(B-V)$ is not much smaller (and $A_{\rm v}$ is larger) for Herschel 36; a number of others (not listed) are not detected at all toward Herschel 36.
Only one DIB, at 6234.0 \AA, appears to be significantly stronger toward Herschel 36 than toward both HD 183143 and HD 204827.
The last two columns in the table indicate the presence or absence of extended redward wings in the DIBs (as seen in the ARCES and FEROS spectra of Herschel 36) and note potential blends with other DIBs.  
Besides the five DIBs already discussed, seven other DIBs ($\lambda\lambda$4727.2, 4762.6, 5849.8, 6010.7, 6234.0, 6376.2, and 6379.3) may also exhibit redward wings.
Seven cases are ambiguous, with slight differences between the ARCES and FEROS spectra; most of these are for relatively weak DIBs, where we are limited by the S/N in searching for wings.
 
The complex feature at 6203.1--6204.8 \AA\ is less clearly related to the redward wings seen for other DIBs, since in this case the ``wing'' appears in the mean spectrum for the full sample of O and B stars (black trace in Figure~\ref{fig:meandib}, lower right) -- i.e., not just for the few stars showing evidence of redward wings at $\lambda$5780.5 and $\lambda$6196.0 in Figure~\ref{fig:meandib} (red, purple, green and yellow).  
Those latter few stars do, however, show enhanced absorption over the mean (lower right panel of Figure~\ref{fig:meandib}).  
Perhaps there are two persistent DIBs -- $\lambda$6203.1 and $\lambda$6204.8 -- where the DIB at 6203.1 \AA\ can have a redward wing, coincidentally blended with a second persistent DIB near 6204.8 \AA.  
[We use the average of the slightly different wavelengths for the rightmost DIB found in Hobbs et al. (2008, 2009).]  
In studies which included the Trapezium stars and HD 37903, Benvenuti \& Porceddu (1989) and Porceddu et al. (1991) concluded that the feature should be considered as two separate DIBs.  

\begin{deluxetable}{lrrrrll}
\tablecolumns{7}
\tabletypesize{\scriptsize}
\tablecaption{Strengths and Profiles of DIBs: Herschel 36 versus HD 183143 and HD 204827. \label{tab:3los}}
\tablewidth{0pt}

\tablehead{
\multicolumn{1}{c}{DIB}&
\multicolumn{3}{c}{Equivalent Widths (m\AA)}&
\multicolumn{1}{c}{FWHM}&
\multicolumn{1}{c}{Profile}&
\multicolumn{1}{c}{Blend?}\\
\multicolumn{1}{c}{ } &
\multicolumn{1}{c}{Herschel 36}&
\multicolumn{1}{c}{HD 183143}&
\multicolumn{1}{c}{HD 204827}&
\multicolumn{1}{c}{(\AA)}&
\multicolumn{1}{c}{ }&
\multicolumn{1}{c}{ }}

\startdata
4501.8                  & 118.8 & 211.2 &   31.5 &  2.53 & Ambiguous & \nodata   \\ 
4726.8\tablenotemark{a} & 120.9 & 156.2 &  283.8 &  2.93 & Wing?     & \nodata   \\ 
4762.6                  &  64.7 & 126.5 &   13.7 &  1.75 & Wing?     & \nodata   \\ 
4963.9\tablenotemark{a} &  11.5 &  26.4 &   53.4 &  0.67 & No Wing   & \nodata   \\ 
4984.8\tablenotemark{a} &   6.3 &  11.5 &   31.1 &  0.56 & No Wing   & \nodata   \\ 
5176.0\tablenotemark{a} &   2.8 &   3.0 &   35.6 &  0.67 & \nodata   & \nodata   \\ 
5363.7                  &   5.0 &  29.8 &   10.8 &  1.37 & No Wing   & \nodata   \\ 
5404.6                  &  13.1 &  52.9 &   14.9 &  1.01 & Ambiguous & \nodata   \\ 
5418.9\tablenotemark{a} &   3.1 &  13.1 &   49.1 &  0.80 & No Wing   & \nodata   \\ 
5494.1                  &  10.5 &  31.2 &   24.2 &  0.60 & No Wing   & 5497.1    \\ 
5508.1                  &  55.8 & 158.8 &   84.1 &  2.52 & No Wing   & \nodata   \\ 
5512.7\tablenotemark{a} &   5.5 &   8.2 &   20.8 &  0.50 & No Wing   & \nodata   \\ 
5705.1                  &  79.7 & 172.5 &   41.6 &  2.63 & No Wing?  & 5707.8    \\ 
5719.5                  &   9.1 &  21.7 &   16.7 &  0.81 & No Wing   & \nodata   \\ 
5780.5                  & 446.5 & 779.3 &  257.0 &  2.13 & Wing      & \nodata   \\ 
5797.1                  & 109.4 & 186.4 &  199.0 &  0.84 & Wing      & \nodata   \\ 
5849.8                  &  39.2 &  67.8 &   95.6 &  0.88 & Wing      & \nodata   \\ 
5923.5                  &   5.1 &  30.4 &   13.9 &  0.71 & No Wing   & 5925.9    \\ 
6010.8                  &  33.1 & 202.8 &   31.5 &  3.64 & Wing?     & Blend?    \\ 
6027.7                  &  15.2 &  57.8 &   18.5 &  2.04 & Ambiguous & 6030.5    \\ 
6037.6                  &  13.5 &  77.3 &   14.8 &  2.80 & No Wing   & \nodata   \\ 
6089.9                  &  10.5 &  23.7 &   28.0 &  0.59 & No Wing   & \nodata   \\ 
6113.2                  &   5.2 &  41.5 &   24.3 &  0.80 & No Wing   & 6116.8    \\ 
6196.0                  &  31.2 &  90.4 &   37.8 &  0.54 & Wing      & \nodata   \\ 
6203.1                  & 146.2 & 206.2 &   57.1 &  1.32 & Wing?     & 6204.8    \\ 
6234.0                  &  54.2 &  18.9 &   25.3 &  0.72 & Wing      & 6236.9    \\ 
6269.9                  & 129.8 & 256.4 &   77.0 &  1.25 & No Wing   & \nodata   \\ 
6376.1\tablenotemark{b} & 100.5 &  63.7 &   44.6 &  0.85 & Wing      & 6379.3    \\ 
6379.3\tablenotemark{b} & 100.5 & 105.4 &   94.9 &  0.63 & Wing      & Stellar?  \\ 
6397.0                  &  13.9 &  26.4 &   25.0 &  1.12 & Ambiguous & 6400.5    \\ 
6425.7                  &   5.7 &  26.4 &    5.7 &  0.73 & No Wing   & \nodata   \\ 
6439.5                  &   5.0 &  26.9 &   25.4 &  0.75 & No Wing   & \nodata   \\ 
6445.3                  &   9.4 &  58.1 &   35.7 &  0.65 & Ambiguous & \nodata   \\ 
6613.6                  & 131.2 & 341.6 &  165.1 &  1.01 & Wing      & \nodata   \\ 
6633.1                  &   5.3 &  21.6 &    9.7 &  1.22 & \nodata   & \nodata   \\ 
6660.7                  &   6.9 &  59.7 &   33.0 &  0.63 & No Wing   & 6662.2    \\ 
6699.3                  &  12.7 &  43.3 &   21.6 &  0.73 & Ambiguous & 6702.0    \\ 
6740.8                  &   2.4 &  20.7 &    3.3 &  1.03 & \nodata   & \nodata   \\ 
6770.2                  &   2.1 &  22.1 &    4.9 &  0.71 & \nodata   & \nodata   \\ 
6843.6                  &   6.2 &  46.4 &   12.3 &  1.07 & No Wing   & 6845.3    \\ 
6944.6                  &   6.4 &  30.8 &   10.8 &  0.84 & Ambiguous & \nodata   \\ 
7224.0                  &  70.3 & 358.8 &   84.9 &  1.13 & No Wing   & \nodata   \\ 
7334.5                  &  10.4 &  60.6 &    9.8 &  0.89 & No Wing   & \nodata   \\ 
7357.6                  &  10.2 &  53.4 &   12.5 &  1.00 & No Wing   & 7360.6    \\ 
7562.3                  &  21.8 & 100.5 &   29.2 &  1.48 & No Wing   & 7564.5    \\ 
\enddata
\tablenotetext{a}{C$_2$ DIB (Thorburn et al. 2003)}
\tablenotetext{b}{Total equivalent width for blended feature given for Herschel 36}
\tablecomments{For DIBs between 4000 and 8000 \AA.  FWHM are average values for HD 183143 and HD 204827 (Hobbs et al. 2008, 2009); DIBs with FWHM $>$ 4 \AA\ are not included.}
\end{deluxetable}

\noindent
Hobbs et al. (2008, 2009) also listed them as two features ($\lambda\lambda$6203.1, 6205.2 toward HD 183143; $\lambda\lambda$6203.1, 6204.5 toward HD 204827).  
Friedman et al. (2011), however, measured the apparent blend as a single DIB and found a very good correlation of the total strength of the complex feature with that of the $\lambda$5780.5 DIB.  
We conclude that the conditions that lead to extended redward wings for the $\lambda$5780.5, $\lambda$6196.0, and $\lambda$6613.6 DIBs either lead to an augmentation of a pervasive wing-like feature on the DIB at 6203.1 \AA\ (namely, the feature near 6204.8 \AA\ that occurs toward all stars) or that an independent, persistent feature near 6204.8 \AA\ is separately enhanced in the stars showing evidence of extended redward wings for the main DIBs in this study.  
Note that among the stars in the bottom right panel of Figure~\ref{fig:meandib}, the spectrum of HD 37061 has the most anomalous strength of the $\lambda$6204.8 DIB.

We have also examined the ARCES and FEROS spectra of Herschel 36 to look for new DIBs that are not present in our high-S/N atlas spectra of HD 183143 or HD 204827 (Hobbs et al. 2008, 2009).
Over the wavelength range from about 4000 to 8000 \AA, there do not appear to be any new DIBs stronger than about 10 m\AA, though a weak feature, with $W$ $\sim$ 3--4 m\AA, may be present at about 5165.0 \AA.
It thus appears that the extreme UV radiation field (due to the proximity of Herschel 36) is not contributing to the creation of easily detectable new DIBs, for instance, by ionizing or dissociating some carriers and producing new carriers with different spectral signatures.  
If such conversion is occurring, then either the DIB signatures of the ionized versions are already in our lists of DIBs, or the new DIBs are weaker than our detection limit, or the new DIBs are to be found outside the wavelength range covered by our spectra.  
 
\subsection{Interstellar Diatomic Molecules toward Herschel 36}
\label{sec-mol}

Of the diatomic molecules commonly seen in optical spectra of reddened stars, only CH$^+$ and CH show relatively strong absorption toward Herschel 36; CN and C$_2$ are not detected (though the strongest lines of CN, near 3875 \AA, are in a region of lower S/N in the currently available spectra).  
The most unusual interstellar absorption lines detected are those from excited rotational levels --  the $J=1$ state of CH$^+$ (E/k = 40.1 K above the ground state), seen for the A-X (0,0) band near 4232 \AA\ and the A-X (1,0) band near 3957 \AA, and the $J=3/2$ state of CH (E/k = 25.6 K above the ground state), seen for the A-X (0,0) band near 4300 \AA\ -- which have not been seen in any other sight line.  
Figure~\ref{fig:chp} shows the ARCES spectrum of the CH$^+$ A-X (0,0) band (in which the excited lines were initially seen); higher S/N FEROS spectra of CH$^+$ and CH (which confirmed the tentative detection of the excited CH line in the ARCES spectrum) are shown in Fig.~\ref{fig:atmdibs}.  
In Paper II, column densities for CH$^+$ and CH toward Herschel 36 are derived by requiring consistent fits to the profiles of lines of different strength available in the optical spectra.
Ratios of the column densities in the excited and ground states, $N$($J=1$)/$N$($J=0$) for CH$^+$ and $N$($J=3/2$)/$N$($J=1/2$) for CH, imply excitation temperatures of 14.6 and 6.7 K for CH$^+$ and CH, respectively.   
The excitation is likely due to strong, local infrared radiation from the adjacent source Her 36 SE (Paper II). 

Absorption from H$_2$ has been detected in far-UV spectra of Herschel 36 obtained with {\it FUSE} (B. Rachford et al., in preparation).
In addition to lines from all 14 rotational levels of the ground vibrational state, more than 100 lines from various rotational levels of the first four vibrationally excited states of molecular hydrogen have been identified in the {\it FUSE} spectrum.
Such absorption from vibrationally excited interstellar H$_2$ has been seen in only a few other cases (Meyer et al. 2001; Boiss\'{e} et al. 2005).  
The strength of the excited H$_2$ toward Herschel 36 appears to be broadly comparable to that seen toward HD 37903 [the illuminating star of the reflection nebula NGC 2023 in the Orion Molecular Cloud (Meyer et al. 2001) and one of the stars exhibiting slightly extended redward wings on several of the DIBs in Fig.~\ref{fig:meandib}]; the total column density of excited H$_2$ is less than 10$^{19}$ cm$^{-2}$ (B. Rachford et al., in preparation).  
The many H$_2$ lines detected in the {\it HST} and {\it FUSE} spectra of HD 37903 have been modeled and interpreted as the result of relatively dense molecular gas within 1 pc of that B1.5 V star, undergoing fluorescent excitation due to the intense UV radiation field (Meyer et al. 2001; Gnaci\'{n}ski 2011).  
Herschel 36 is within 0.1 pc of the Hourglass and even closer to the dark lane (Figure~\ref{fig:prompt}), so it is not surprising to see the vibrationally excited H$_2$ there, despite the high extinction.
The excited H$_2$ is not detected in {\it FUSE} spectra of HD 164816 or HD 164906 (two other stars in NGC 6530), at projected separations of 2.4 and 4.6 pc from Herschel 36, respectively (B. Rachford et al., in preparation).

So far, we have not found absorption from excited CH$^+$ in any other sight line -- either in our database of high S/N ARCES spectra or in spectra retrieved from the ESO archive, originally obtained (in many cases) to search for weak molecular absorption lines (e.g., Roueff et al. 2002; Casassus et al. 2005).
To our knowledge, the only other detection of excited CH$^+$ in absorption is from an inferred circumbinary disk around the post-AGB star HD 213985 (Bakker et al. 1997).
Emission from excited CH$^+$ and CH has been detected in the planetary nebula NGC 7027 (Cernicharo et al. 1997; Black 1998) and toward HD 44179, the illuminating star of the Red Rectangle (Balm \& Jura 1992; Hobbs et al. 2004).
More recently, {\it Herschel} spectra of the CH$^+$ $J=1-0$ transition at 835 GHz have revealed emission from the star-forming region DR21 (Falgarone et al. 2010) and from the Orion bar (Naylor et al. 2010; Nagy et al. 2013).
The lines seen in several of those cases involve even higher $J$ values than those seen in absorption toward Herschel 36.  
It has been suggested that some DIBs, most notably $\lambda$5797.1, are also seen in emission from the Red Rectangle (Sarre 1991; Scarrott et al. 1992; Sharp et al. 2006; Wehres et al. 2011). 
There thus might be some connection between the appearance of those DIBs in emission and the presence of excited CH$^+$.
At the suggestion of A. Witt, we have tried to detect an emission feature near the $\lambda$5797.1 DIB in spectra of the Hourglass Nebula near Herschel 36, but have not yet succeeded.

\subsection{Observations of Neighboring Sight Lines}
\label{sec-neigh}

In order to probe the interstellar material in the region around Herschel 36, we have obtained optical spectra of eight other stars in the vicinity (within 2 degrees), either performing new observations (with ARCES or MIKE) or retrieving existing spectra from the ESO archive (FEROS or UVES).  
Some data for DIBs, H, and/or H$_2$ were obtained from the literature or (for H) derived from archival {\it IUE} spectra for nine other sight lines.  
Table~\ref{tab:region} lists the various sight lines, with Galactic coordinates; angular separation from Herschel 36; distance; $E(B-V)$; column densities of atomic hydrogen, H$_2$, CH, and CH$^+$; equivalent widths of the DIBs at 5780.5, 5797.1, 6196.0, and 6613.6 \AA; and the sources of the optical spectra.
For the neighboring sight lines, even the strongest lines of CH ($\lambda$4300.3) and CH$^+$ ($\lambda$4232.5) in most cases have equivalent widths less than 10 m\AA, so determination of column densities is straightforward (via integration over the apparent optical depths of the profiles and/or fits to the profiles).
There is generally good agreement for the equivalent widths of the DIBs measured in spectra obtained with different instruments.
Most of the stars are likely members of NGC 6530, at an assumed distance of 1500 pc, with separations from Herschel 36 ranging from 1.3 to 80 arcmin.  
HD 165814 and HD 164402 are the only foreground targets, at distances of about 580 and 1250 pc and separations of about 97 and 99 arcmin (respectively) from Herschel 36.  

\begin{deluxetable}{lrrrrrrrrrrrrrl}
\rotate
\tabletypesize{\scriptsize}
\tablecolumns{15}
\tablecaption{Interstellar gas, dust, and DIBs toward Herschel 36 and 17 other stars \label{tab:region}}
\tablewidth{0pt}

\tablehead{
\multicolumn{1}{l}{Star} &
\multicolumn{1}{c}{l} &
\multicolumn{1}{c}{b} &
\multicolumn{1}{c}{sep} &
\multicolumn{1}{c}{dist} &
\multicolumn{1}{c}{$E(B-V)$} &
\multicolumn{1}{c}{$N$(H)} &
\multicolumn{1}{c}{$N$(H$_2$)} &
\multicolumn{1}{c}{$N$(CH)} &
\multicolumn{1}{c}{$N$(CH$^+$)} &
\multicolumn{1}{c}{5780} &
\multicolumn{1}{c}{5797} &
\multicolumn{1}{c}{6196} &
\multicolumn{1}{c}{6613} &
\multicolumn{1}{c}{Instr} \\
\multicolumn{1}{c}{ } &
\multicolumn{1}{c}{ } &
\multicolumn{1}{c}{ } &
\multicolumn{1}{c}{(')} &
\multicolumn{1}{c}{(pc)} &
\multicolumn{1}{c}{ } &
\multicolumn{1}{c}{ } &
\multicolumn{1}{c}{ } &
\multicolumn{1}{c}{ } &
\multicolumn{1}{c}{ } &
\multicolumn{1}{c}{(m\AA)} &
\multicolumn{1}{c}{(m\AA)} &
\multicolumn{1}{c}{(m\AA)} &
\multicolumn{1}{c}{(m\AA)} &
\multicolumn{1}{c}{ }} 

\startdata
Herschel 36   & 5.97 & $-$1.17 &  0.0 &  1500 &  0.87 & 21.95$\pm$0.15 & 20.19$\pm$0.12 & 13.87$\pm$0.03 & 14.04$\pm$0.04 
                                                      &   440$\pm$ 5   &   110$\pm$ 2   &    32$\pm$ 1   &   131$\pm$ 3   & FEROS \\
              &      &         &      &\nodata&\nodata&\nodata         &\nodata         &$>$13.64        &$>$13.84
                                                      &   463$\pm$ 8   &   102$\pm$ 7   &    28$\pm$ 2   &   132$\pm$ 6   & ARCES \\
              &      &         &      &\nodata&\nodata&\nodata         &\nodata         & 13.78$\pm$0.04 & 13.96$\pm$0.04
                                                      &   463$\pm$ 7   &   117$\pm$ 3   &    30$\pm$ 2   &   133$\pm$ 3   & MIKE \\
 & \\
CD$-$24 13810 & 5.99 & $-$1.18 &  1.3 &  1500 &  0.38 &\nodata         &\nodata         & 12.90$\pm$0.11 & 12.96$\pm$0.06 
                                                      &   166$\pm$ 9   &\nodata         &    18$\pm$ 2   &    67$\pm$ 3   & ARCES \\
              &      &         &      &\nodata&\nodata&\nodata         &\nodata         & 12.83$\pm$0.02 & 12.96$\pm$0.02 
                                                      &   175$\pm$ 4   &    44$\pm$ 1   &    17$\pm$ 1   &    62$\pm$ 1   & MIKE \\
N6530 CDZ98   & 5.99 & $-$1.19 &  1.6 &  1500 &\nodata&\nodata         &\nodata         & 12.85$\pm$0.07 & 12.70$\pm$0.10 
                                                      &   167$\pm$16   &    33$\pm$ 4   &    19$\pm$ 3   &    46$\pm$ 5   & MIKE \\
9 Sgr         & 6.01 & $-$1.21 &  3.0 &  1500 &  0.34 & 21.29$\pm$0.07 & 20.10$\pm$0.06 & 12.93$\pm$0.02 & 13.06$\pm$0.01 
                                                      &   142$\pm$ 1   &    35$\pm$ 1   &    16$\pm$ 1   &    52$\pm$ 1   & FEROS \\
              &      &         &      &\nodata&\nodata&\nodata         &\nodata         & 12.94$\pm$0.02 & 13.04$\pm$0.02 
                                                      &   154$\pm$ 2   &    40$\pm$ 1   &    15$\pm$ 1   &    49$\pm$ 2   & UVES \\ 
HD 164816     & 6.06 & $-$1.20 &  5.5 &  1500 &  0.30 & 21.18$\pm$0.13 & 20.03$\pm$0.04 & 12.88$\pm$0.02 & 12.85$\pm$0.03 
                                                      &   129$\pm$ 5   &    33$\pm$ 1   &    18$\pm$ 1   &    56$\pm$ 3   & FEROS \\
 & \\
near neighbor &      &         &      &  1500 &  0.34 & 21.24$\pm$0.08 & 20.07$\pm$0.05 & 12.88$\pm$0.04 & 12.89$\pm$0.15 
                                                      &   154$\pm$19   &    37$\pm$ 5   &    18$\pm$ 1   &    54$\pm$ 8   & avg \\
 & \\
HD 315032     & 6.02 & $-$1.29 &  7.9 &  1500 &  0.28 & 21.26$\pm$0.08 &\nodata         &\nodata&\nodata&\nodata&\nodata&\nodata&\nodata&\nodata\\
CD$-$24 13829 & 6.07 & $-$1.30 &  9.4 &  1500 &  0.36 & 21.32$\pm$0.13 &\nodata         &\nodata&\nodata&\nodata&\nodata&\nodata&\nodata&\nodata\\
HD 315033     & 6.00 & $-$1.34 & 10.4 &  1500 &  0.35 & 21.20$\pm$0.09 &\nodata         &\nodata&\nodata&\nodata&\nodata&\nodata&\nodata&\nodata\\
HD 164906     & 6.05 & $-$1.33 & 10.4 &  1500 &  0.43 & 21.20$\pm$0.09 & 20.23$\pm$0.04 &\nodata&\nodata&\nodata&\nodata&\nodata&\nodata&\nodata\\
HD 315026     & 6.15 & $-$1.22 & 11.2 &  1500 &  0.32 & 21.34$\pm$0.10 &\nodata         &\nodata&\nodata&\nodata&\nodata&\nodata&\nodata&\nodata\\
HD 315021     & 6.12 & $-$1.33 & 13.0 &  1500 &  0.34 & 21.28$\pm$0.10 & 19.99$\pm$0.02 &\nodata&\nodata&\nodata&\nodata&\nodata&\nodata&\nodata\\
CD$-$24 13785 & 5.97 & $-$0.91 & 15.9 &  1500 &  0.28 & 21.15$\pm$0.13 &\nodata         &\nodata&\nodata&\nodata&\nodata&\nodata&\nodata&\nodata\\
HD 164536     & 5.96 & $-$0.91 & 15.9 &  1500 &  0.25 & 21.18$\pm$0.06 &\nodata         &\nodata&\nodata&\nodata&\nodata&\nodata&\nodata&\nodata\\
HD 165052     & 6.12 & $-$1.48 & 20.6 &  1500 &  0.42 & 21.36$\pm$0.10 & 20.20$\pm$0.xx & 12.90$\pm$0.03 & 13.40$\pm$0.01 
                                                      &   186$\pm$ 3   &    46$\pm$ 1   &    20$\pm$ 1   &    69$\pm$ 2   & FEROS \\
HD 165246     & 6.40 & $-$1.56 & 34.7 &  1500 &  0.40 & 21.41$\pm$0.07 & 20.15$\pm$0.08 & 12.81$\pm$0.01 & 13.05$\pm$0.01 
                                                      &   174$\pm$ 2   &    40$\pm$ 1   &    20$\pm$ 1   &    65$\pm$ 2   & FEROS \\
HD 165921     & 6.94 & $-$2.10 & 80.4 &  1500 &  0.46 &\nodata         &\nodata         & 12.93$\pm$0.02 & 12.97$\pm$0.02 
                                                      &   283$\pm$ 4   &    68$\pm$ 1   &    25$\pm$ 1   &    91$\pm$ 2   & FEROS \\
 & \\
all neighbor  &      &         &      &  1500 &  0.35 & 21.27$\pm$0.08 & 20.12$\pm$0.09 & 12.88$\pm$0.04 & 13.00$\pm$0.22 
                                                      &   180$\pm$49   &    43$\pm$12   &    19$\pm$ 3   &    63$\pm$15   & avg \\
 & \\
HD 165814     & 5.59 & $-$2.74 & 96.8 &   580 &  0.22 &\nodata         &\nodata         & 12.59$\pm$0.02 & 12.92$\pm$0.02 
                                                      &   117$\pm$ 4   &    27$\pm$ 1   &    12$\pm$ 1   &    34$\pm$ 1   & MIKE \\
HD 164402     & 7.16 & $-$0.03 & 98.9 &  1250 &  0.24 & 21.30$\pm$0.13 & 19.49$\pm$0.14 &\nodata         &\nodata
                                                      &   187$\pm$19   &    52$\pm$ 7   &\nodata         &\nodata         & Lick  \\
\enddata
\tablecomments{References:  
Column densities of H and H$_2$ are from Savage et al. 1977; Fitzpatrick \& Massa 1990, 2007; Diplas \& Savage 1994; Snow et al. 2007; Cartledge et al. 2004, 2008; Rachford et al. 2009; B. Rachford et al., in preparation; or this work.  
Column densities for CH and CH$^+$ are from this work or Paper II.
Equivalent widths for the DIBs are from Herbig 1993 (HD 164402), Weselak et al. 2008 (9 Sgr: second $\lambda$5780, $\lambda$5797), Friedman et al. 2011 (ARCES Herschel 36), or this work.}
\end{deluxetable}

\begin{figure}[b!]
\plottwo{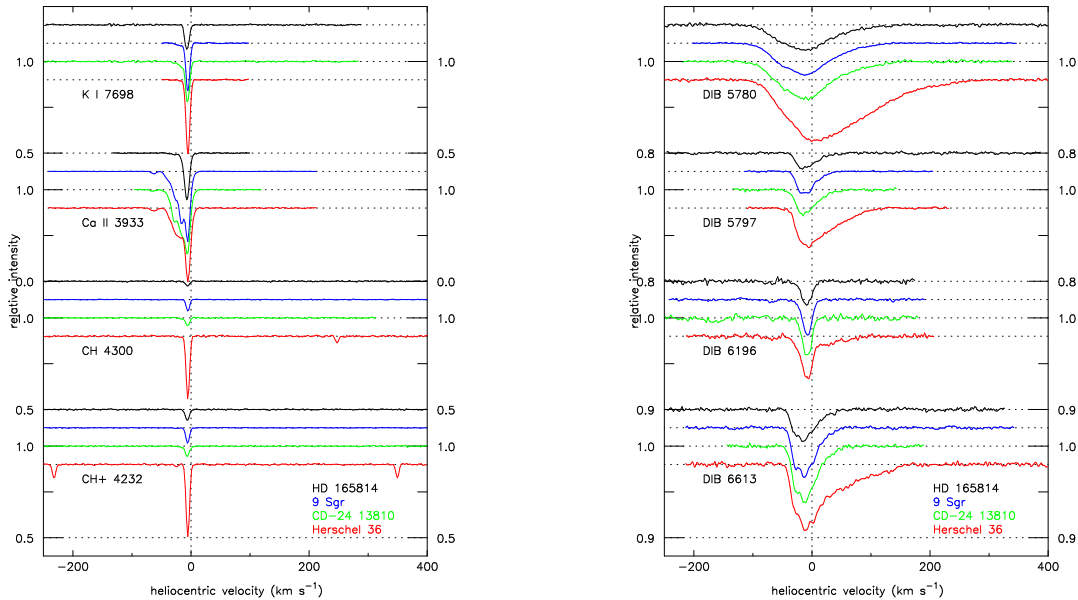}{f7b.eps}
\caption{Profiles for interstellar \ion{K}{1} $\lambda$7698, \ion{Ca}{2} $\lambda$3933, CH $\lambda$4300, and CH$^+$ $\lambda$4232 absorption (left) and for the $\lambda\lambda$5780.5, 5797.1, 6196.0, and 6613.6 DIBs (right), for Herschel 36 and three other stars.
The spectra are from FEROS or MIKE, at resolutions $\sim$ 50,000.  
The stars CD$-$24 13810 and 9 Sgr (also members of NGC 6530) are both within 3 arcmin of Herschel 36 and at similar distances ($\sim$1500 pc); HD 165814 is in the foreground (at $\sim$580 pc), $\sim$1.6 degrees from Herschel 36.  
In all four sight lines, all four species exhibit a dominant component near $v_{\odot}$ $\sim$ $-$5.5 km s$^{-1}$ (much stronger toward Herschel 36); several weaker components at more negative velocities are seen in \ion{Ca}{2} toward the three stars in NGC 6530.  
For Herschel 36, the weak features at higher velocities for CH and CH$^+$ are the lines from rotationally excited levels; the very weak feature for CH$^+$ at $v_{\odot}$ $\sim$ $-$23 km s$^{-1}$ is due to $^{13}$CH$^+$.  
The four DIBs exhibit extended wings to the red only toward Herschel 36.  
The DIB profiles for the other two stars in NGC 6530 and for the foreground star HD 165814 resemble those typically seen in the local Galactic ISM.}
\label{fig:atmdibs}
\end{figure}  
 
The profiles of several atomic and molecular lines and of the four DIBs of most interest here, for Herschel 36 and three of the other nearby sight lines, are shown in Figure~\ref{fig:atmdibs}.  
The left-hand panel shows the spectra of \ion{K}{1} $\lambda$7698, \ion{Ca}{2} $\lambda$3933, CH$^+$ $\lambda$4232, and CH $\lambda$4300 for Herschel 36, CD$-$24 13810, and 9 Sgr (stars 1, 2, and 3 in Fig.~\ref{fig:prompt}; all members of NGC 6530), as well as for the foreground star HD 165814.  
Both CD$-$24 13810 and 9 Sgr are within 3 arcmin (1.3 pc at 1500 pc) of Herschel 36; HD 165814 is about 1.6 degree (16 pc at 580 pc) away.  
The right-hand panel shows the profiles of the four DIBs ($\lambda\lambda$5780.5, 5797.1, 6196.0 and 6613.6).  
The spectra of Herschel 36 and 9 Sgr were obtained with FEROS; the spectra of HD 165814 and CD$-$24 13810 were obtained with MIKE.  
All of the sight lines near Herschel 36 with optical spectra (including the foreground) are relatively simple in \ion{K}{1}, CH, and CH$^+$ -- with a single dominant component (at these resolutions) near $v_{\odot}$ $\sim$ $-$5.5 km s$^{-1}$.  
That main component is much stronger toward Herschel 36, for all three of those species.  
For the stars in NGC 6530, the profiles of \ion{Ca}{2} and \ion{Na}{1} $\lambda\lambda$5889, 5895 also exhibit several weaker components at more negative velocities.  
Those weaker components have similar spreads and component structures across the region, and presumably represent some common foreground material, perhaps associated with M8 but not specific to the Hourglass region.  
For 9 Sgr, where higher resolution UVES spectra are available, only the two highest velocity components, at $-$63 and $-$40 km s$^{-1}$, exhibit the very low $N$(\ion{Na}{1})/$N$(\ion{Ca}{2}) ratios ($\la$ 0.2) characteristic of the Routly-Spitzer effect (Routly \& Spitzer 1952); the other weak components have $N$(\ion{Na}{1})/$N$(\ion{Ca}{2}) $\sim$ 1.
The molecules CN and C$_2$, the least ubiquitous diatomic molecules in our DIB survey, are not detected toward Herschel 36, and are either weak (CN toward 9 Sgr) or undetected in the various neighboring sight lines listed in Table~\ref{tab:region}.
The profiles of the four DIBs toward Herschel 36 all exhibit extended redward wings -- and thus differ markedly from the corresponding profiles toward the other three stars, which resemble those seen in the local Galactic ISM (e.g., Fig.~\ref{fig:meandib}).  
In particular, no extended redward wings are seen in the spectra of the other stars in NGC 6530.  

\section{DISCUSSION} 
\label{sec-disc}

\subsection{Location and Properties of the Material Producing the Extended DIB Wings}
\label{sec-loc}

Toward Herschel 36, the unique combination of absorption from rotationally excited CH$^+$ and CH, the anomalous extinction curve, the unusual broad wings on several DIBs (Figure~\ref{fig:4dibs}), and the proximity of the bright infrared source Her 36 SE suggests that the excited levels of CH$^+$ and CH are pumped by the infrared photons from Her 36 SE and that the extended, redward wings on several DIBs are caused by pumping of the rotational levels in the ground states of the DIB carrier molecules (Paper II).
The presence of those unusual spectral features toward Herschel 36, their absence toward other stars in NGC 6530, and the morphological features noted in Figures~\ref{fig:prompt} and \ref{fig:goto} together suggest that both the excited molecular species and the DIB carriers producing the extended redward wings on the DIBs in this sight line are located in a relatively small (``local'') region near Herschel 36 and Her 36 SE.  
None of those unusual features is seen toward the nearest neighbor, CD$-$24~13810 (which is only 1.3 arcmin, or 0.6 pc at 1500 pc) away from Herschel 36.
The dark lane in front of Herschel 36, visible in Figure~\ref{fig:prompt} and likely responsible for the much higher $E(B-V)$ and $A_{\rm v}$ toward Herschel 36 (\S~\ref{sec-region}), appears to be only $\sim$4000 AU across, if at the distance of Herschel 36.
In order to determine the properties of the small region local to Herschel 36, however, we must identify the contributions to the observed spectral features from any unrelated material along the line of sight.

While the properties of the neighboring NGC 6530 sight lines appear to be fairly uniform (with averages given in the ``near neighbor'' and ``all neighbor'' lines in Table~\ref{tab:region}), some variations may be noted in $E(B-V)$, the column densities, and the DIB equivalent widths.
Those variations may be due to small-scale structure within the M8 region and/or to differences in depth of the various target stars within that region.
The properties of the neighboring sight lines differ significantly in several respects from those toward Herschel 36, however.  
The color excess of Herschel 36 is 0.87, compared to the average value 0.35$\pm$0.06 for the neighboring sight lines.
The far-UV extinction is much flatter, the 2175 \AA\ bump is weaker, and $R_{\rm v}$ is significantly larger toward Herschel 36 (\S~\ref{sec-region}). 
The column density of atomic hydrogen toward Herschel 36, log $N$(H) = 21.95, is a factor of 5 higher than the mean value for the eight neighboring sight lines in Table~\ref{tab:region} with measured values:  log $N$(H) = 21.27$\pm$0.08.
The column density of molecular hydrogen toward Herschel 36, log $N$(H$_2$) = 20.19, however, is consistent with the average for the six neighboring sight lines with data:  log $N$(H$_2$) = 20.12$\pm$0.09.  
The column densities of CH and CH$^+$ toward Herschel 36 exceed the average values for the seven neighboring sight lines with data by factors of $\sim$10 (much larger than the spread in the values for the neighbors), and vibrationally excited H$_2$ is detected only toward Herschel 36.
The four DIBs pictured in Figures~\ref{fig:4dibs} and \ref{fig:atmdibs} are roughly twice as strong toward Herschel 36 as the averages over the neighboring sight lines (which have rms scatter less than 30 percent).  

\subsubsection{A Known Foreground Cloud}
\label{sec-rc}
 
Given the roughly 1500 pc distance to Herschel 36 (and the rest of NGC 6530), there may well be significant contributions to the observed interstellar absorption from foreground material unrelated to the cluster.  
One such concentration of intervening material is known from the work of Riegel \& Crutcher (1972) and Crutcher \& Lien (1984), who discussed \ion{H}{1} self-absorption maps (derived from 21 cm spectra) and optical \ion{Na}{1} absorption lines, respectively, in a region larger than 200 square degrees that overlaps M8 and the region of interest for this paper.  
They identified self-absorption in the 21 cm line over a fairly broad region and strong \ion{Na}{1} absorption toward a number of the more distant stars in their sample, which they attributed to an extensive sheet-like body of gas at a distance of 125$\pm$25 pc (or $\sim$150 pc, using more recent distances from {\it Hipparcos}) from the Sun in that large region.  
Higher resolution images of the \ion{H}{1} self-absorption have revealed a complex network of narrow filaments, aligned with the local magnetic field, in one large section of this cloud (McClure-Griffiths et al. 2006).
Stars located immediately behind the Riegel--Crutcher (R--C) cloud (at distances of 160--400 pc) within 2 degrees of Herschel 36 have $E(B-V)$ ranging from about 0.14 to 0.25 (P. Frisch, private communication).
There appears to be comparatively little interstellar material at smaller distances in this direction.

Among the stars in Crutcher \& Lien's sample exhibiting strong \ion{Na}{1} absorption is HD 165814, a B3 III star located 1.6 degrees from Herschel 36 and about 580 pc from the Sun.  
Column densities of CH and CH$^+$ and DIB equivalent widths derived from our MIKE spectrum of HD 165814 are included in Table~\ref{tab:region}; the corresponding line profiles are shown in Fig.~\ref{fig:atmdibs}.  
The measured quantities are similar to the values for the other neighboring sight lines in the table:  the color excess (0.22), $N$(CH), $N$(CH$^+$), and equivalent widths of the DIBs are at (or slightly below) the low end of the range seen for the more distant targets.  
The velocity of the nearby material is very close to that of the main component seen toward the more distant stars (Fig.~\ref{fig:atmdibs}).  
Much of the material (at least half) seen toward the neighboring stars in NGC 6530 thus may be associated with this foreground cloud.  
There could be other, similar clouds or cloud complexes in the line of sight, but there is no indication of other significant velocity components or significant increments in the strengths of the lines of H, H$_2$, or DIBs between the R--C cloud and NGC 6530.

\subsubsection{Estimating the Total Foreground Contribution}
\label{sec-fore}

The total foreground interstellar contribution for the Herschel 36 sight line is given primarily by the Riegel--Crutcher cloud (at $\sim$ 150 pc) and any more distant clouds (perhaps including some associated with M8 and/or NGC 6530) that lie between the R--C cloud and the local material very near Herschel 36 that we wish to characterize.
In order to estimate that foreground, we consider the four nearest neighbor sight lines in Table~\ref{tab:region}:  toward CD$-$24~13810, NGC 6530 CDZ 98, 9 Sgr, and HD 164816, which are all in NGC 6530, within 6 arcmin (2.6 pc at 1500 pc) of Herschel 36.
While those four stars all lie to the NE of Herschel 36 (Fig.~\ref{fig:prompt}), the R--C cloud extends beyond Herschel 36 for at least several degrees in all directions (Riegel \& Crutcher 1972).
Of those four, the sight line toward 9 Sgr has the most extensive and highest quality data, and is also quite representative of the other neighboring sight lines in NGC 6530.
The UV extinction curve for 9 Sgr (Patriarchi \& Perinotto 1999) is very similar to the curves for 21 other members of NGC 6530 (excluding Herschel 36; Fitzpatrick \& Massa 2007), the $R_{\rm v}$ = 3.57 for 9 Sgr (Wegner 2003) is consistent with the average of 3.71$\pm$0.28 for that same sample, and the column densities and DIB equivalent widths for 9 Sgr are all within 20 percent of the corresponding mean values for the neighboring NGC 6530 sight lines listed in Table~\ref{tab:region} (which themselves exhibit fairly uniform properties).  
The 3 arcmin angular separation between the sight lines to 9 Sgr and Herschel 36 corresponds to physical separations ranging from 0.13 pc at 150 pc (at the Riegel--Crutcher cloud) -- comparable to the widths of the filaments in that foreground cloud (McClure-Griffiths et al. 2006) -- to 1.3 pc at 1500 pc (at NGC 6530). 
We therefore compare the various measures for the sight lines toward Herschel 36 and 9 Sgr -- assuming that both sample the same foreground material and using the differences to explore the nature of the region very local to Herschel 36.  
Unfortunately, we cannot assign unique velocity components, because the foreground components and the material local to Herschel 36 are blended in both FEROS and ARCES spectra.

\begin{deluxetable}{lrrrrrrr}
\tablecolumns{8}
\tabletypesize{\scriptsize}
\tablecaption{Herschel 36 versus 9 Sgr and Three Stars in Orion \label{tab:local}}
\tablewidth{0pt}

\tablehead{
\multicolumn{1}{l}{Quantity} &
\multicolumn{1}{c}{Herschel 36} &
\multicolumn{1}{c}{9 Sgr} &
\multicolumn{1}{c}{H36 local\tablenotemark{a}} &
\multicolumn{1}{c}{$\theta^1$ Ori B} &
\multicolumn{1}{c}{NU Ori} &
\multicolumn{1}{c}{ } &
\multicolumn{1}{c}{MW avg} \\
\multicolumn{1}{c}{ } &
\multicolumn{1}{c}{ } &
\multicolumn{1}{c}{HD 164794} &
\multicolumn{1}{c}{ } &
\multicolumn{1}{c}{HD 37021} &
\multicolumn{1}{c}{HD 37061} &
\multicolumn{1}{c}{HD 37903} &
\multicolumn{1}{c}{ }}

\startdata
$E(B-V)$                          & 0.87           & 0.34           & 0.53           & 0.54           & 0.52           & 0.35           & \nodata \\
$R_{\rm v}$                       & 5.2-6.0        & 3.57           & (6.1-7.5)      & 5.8-6.4        & 4.55           & 3.95           & 3.1 \\
$A_{\rm v}$                       & 4.5-5.2        & 1.25           & 3.25-3.95      & 3.1-3.5        & 2.37           & 1.38           & \nodata \\
$N$(H)                            & 21.95$\pm$0.15 & 21.29$\pm$0.07 & 21.84$\pm$0.18 & 21.65$\pm$0.13 & 21.78$\pm$0.10 & 21.17$\pm$0.10 & \nodata \\
$N$(H$_2$)                        & 20.19$\pm$0.12 & 20.10$\pm$0.06 &$\la$19.8       & \nodata        & \nodata        & 20.92$\pm$0.06 & \nodata \\
$N$(H$_{\rm tot}$)                & 21.96$\pm$0.15 & 21.34$\pm$0.07 & 21.84$\pm$0.18 &(21.65$\pm$0.13)&(21.78$\pm$0.10)& 21.50$\pm$0.10 & \nodata \\
log $f$(H$_2$)                    & $-$1.47        & $-$0.93        &$\la-$1.7       & \nodata        & \nodata        & $-$0.28        & \nodata \\
 & \\
$N$(K I)                          & 12.08$\pm$0.08 & 11.92$\pm$0.03 & 11.57$\pm$0.09 & 10.68$\pm$0.02 & 10.88$\pm$0.05 & 10.63$\pm$0.04 & \nodata \\
$N$(Na I)                         & $>$12.65       & 13.60$\pm$0.02 & \nodata        & \nodata        & 12.64$\pm$0.05 & 12.62$\pm$0.05 & \nodata \\
$N$(Ca I)                         & 10.26$\pm$0.02 & 10.20$\pm$0.02 &  9.37$\pm$0.11 & \nodata        &  9.76$\pm$0.05 &  9.67$\pm$0.07 & \nodata \\
$N$(Ca II)                        & $>$12.54       & $>$12.59       & \nodata        & \nodata        & \nodata        & 12.14$\pm$0.xx & \nodata \\
 & \\
$N$(CN)                           & $<$11.56       & 11.47$\pm$0.06 & \nodata        & \nodata        &$<$11.67        & 11.85$\pm$0.09 & \nodata \\
$N$(CH)                           & 13.87$\pm$0.03 & 12.93$\pm$0.02 & 13.82$\pm$0.04 &$<$12.46        & 12.13$\pm$0.09 & 12.81$\pm$0.02 & \nodata \\
$N$(CH$^+$)                       & 14.04$\pm$0.04 & 13.06$\pm$0.01 & 13.99$\pm$0.04 & \nodata        & 12.54$\pm$0.10 & 13.08$\pm$0.01 & \nodata \\
 & \\
$W$(5780.5)                       & 453$\pm$5      & 142$\pm$1      & 311$\pm$ 4     &   61$\pm$ 6    & 169$\pm$ 7     & 183$\pm$10     & \nodata \\
$W$(5797.1)                       & 110$\pm$2      &  35$\pm$1      &  75$\pm$ 3     &$<$15           &  35$\pm$ 5     &  33$\pm$ 5     & \nodata \\
$W$(6196.0)                       &  30$\pm$2      &  16$\pm$1      &  14$\pm$ 2     & $<$4           &  13$\pm$ 2     &  12$\pm$ 2     & \nodata \\
$W$(6613.6)                       & 131$\pm$3      &  52$\pm$1      &  79$\pm$ 2     &    6$\pm$ 2    &  34$\pm$ 3     &  36$\pm$ 4     & \nodata \\
 & \\
log K I/(H$_{\rm tot}$)$^2$       & $-$31.84       & $-$30.74       & $-$32.11       & $-$32.62       & $-$32.68       & $-$32.37       & $-$30.61 \\
log CH/H$_2$                      &  $-$6.32       &  $-$7.17       &$\ga-$6.0       & \nodata        & \nodata        &  $-$8.11       &  $-$7.38 \\
log $W$(5780.5)/H                 & $-$19.31       & $-$19.14       & $-$19.37       & $-$19.86       & $-$19.55       & $-$18.91       & $-$18.85 \\
5780.5/5797.1                     &     4.1        &     4.1        &     4.1        &  $>$4.1        &     4.8        &     5.5        & \nodata  \\
log H$_{\rm tot}$/$E(B-V)$        &    22.02       &    21.79       &    22.12       &    21.92       &    22.06       &    21.63       &    21.75 \\
log H$_{\rm tot}$/$A_{\rm v}$     &    21.27       &    21.23       &    21.29       &    21.13       &    21.41       &    21.36       &    21.26 \\
\enddata
\tablenotetext{a}{H36 local = Herschel 36 minus 9 Sgr} 
\tablecomments{References for column densities (for Herschel 36 and 9 Sgr) and DIB equivalent widths (for all stars) are given in Table~\ref{tab:region}. 
Atomic and molecular column densities for the three Orion stars are from ARCES spectra (for weak lines) or from fits to high-resolution spectra (Na I, Ca II).
Values for $R_{\rm v}$ are from Fitzpatrick \& Massa 1990, 2007, 2009; Wegner 2003; or Gordon et al. 2009.  
Average Galactic values for ratios are from Welty \& Hobbs 2001 or Welty et al. 2006.}
\end{deluxetable}

The available interstellar quantities (as well as various ratios that can provide clues to physical conditions) for Herschel 36, 9 Sgr, and the difference between the two sight lines are summarized in Table~\ref{tab:local} (columns 2, 3, and 4).
Given the apparent similarities between the regions around the Orion Trapezium and the M8 Hourglass (\S~\ref{sec-region}) and the redward wings noted for some of the DIBs toward several of the Orion stars (Fig.~\ref{fig:meandib}), corresponding values are also listed for two sight lines in the Trapezium region [with $E(B-V)$ comparable to that for the local material near Herschel 36] and for HD 37903 (exhibiting strong absorption from rotationally excited H$_2$), for comparison (columns 5, 6, and 7).
The last column of the table gives the average values (for the various ratios) found for the local Galactic ISM.  
The quantities listed in Table~\ref{tab:local} include those given in Table~\ref{tab:region}, as well as some derived quantities: the total visual extinction ($A_{\rm v}$, in magnitudes); the total hydrogen column density [$N$(H$_{\rm tot}$) = $N$(H) + 2$N$(H$_2$)]; the fraction of hydrogen in molecular form, $f$ = 2$N$(H$_2$)/$N$(H$_{\rm tot}$); and several ratios.
The $N$(\ion{K}{1})/$N$(H$_{\rm tot}$)$^2$ and $W$(5780)/$N$(H) ratios provide information on the strength of the radiation field; the $N$(CH)/$N$(H$_2$) ratio can be used to characterize the chemistry (thermal versus non-thermal); $N$(H$_{\rm tot}$)/$E(B-V)$ and $N$(H$_{\rm tot}$)/$A_{\rm v}$ are measures of the gas-to-dust ratio.
[We adopt the convention that the column density of atomic hydrogen is denoted by $N$(H), of molecular hydrogen by $N$(H$_2$); and of all hydrogen nuclei in atomic or molecular form by $N$(H$_{\rm tot}$).]
While the local $N$(H$_2$) is formally about (3$\pm$1) $\times$ 10$^{19}$ cm$^{-2}$, the individual values for Herschel 36 and 9 Sgr are within the mutual 1-$\sigma$ uncertainties, so we adopt an upper limit of 6 $\times$ 10$^{19}$ cm$^{-2}$ for the local material.
For the three Orion stars, the atomic and molecular column densities have generally been derived from weak lines in the ARCES spectra or, for \ion{Na}{1} and \ion{Ca}{2}, from fits to high-resolution (FWHM $\la$ 1.5 km s$^{-1}$) spectra of the \ion{Na}{1} D lines ($\lambda\lambda$5889, 5895) or the \ion{Ca}{2} K line ($\lambda$3933) (e.g., Welty \& Hobbs 2001).
The various ratios listed at the bottom of Table~\ref{tab:local} suggest that the properties of the foreground interstellar material [i.e., as seen toward 9 Sgr (column 3)], generally are quite similar to those for the ``Milky Way average'' given in the last column of the table, apart from a somewhat lower $W$(5780.5)/$N$(H) ratio.

\subsubsection{Properties of Material Local to Herschel 36}
\label{sec-local}

The local region near Herschel 36 -- which gives rise to the enhanced (and rotationally excited) CH and CH$^+$, the vibrationally excited H$_2$, the strong redward DIB wings, and the unusual extinction -- must be fairly small.
The various quantities listed for the difference between Herschel 36 and 9 Sgr in column 4 of Table~\ref{tab:local} indicate, however, that that small, local region contains the bulk of both the total hydrogen and the dust seen for the entire line of sight toward Herschel 36, and also permit some characterization of that local region:

\begin{itemize}
\item{As the UV extinction seen for the foreground material toward 9 Sgr is fairly ``normal'', the extinction for the local material near Herschel 36 must be characterized by an even weaker 2175 \AA\ bump and a flatter far-UV than found for the sight line as a whole (e.g., Hecht et al. 1982).
The $R_{\rm v}$ for the local material must be even higher (6.3--7.7) than the already-high value for the entire sight line, suggestive of a significant population of large grains.
The gas-to-dust ratio in the local material, as given by $N$(H$_{\rm tot}$)/$A_{\rm v}$, is not unusual, however.}

\item{While most of the atomic hydrogen is in the local region, most of the H$_2$ appears to be in the foreground material, with a relatively small (and very uncertain) contribution from the local gas.
The molecular fraction $f$(H$_2$) = 2$N$(H$_2$)/$N$(H$_{\rm tot}$) thus appears to be less than about 0.02 in the local region -- which is unusually low, given the high $N$(H$_{\rm tot}$) and $E(B-V)$, for sight lines probed by optical/UV spectroscopy (Savage et al. 1977; Rachford et al. 2009).
Such low molecular fractions may not be atypical for photodissociation regions (PDRs), however.
The local region presumably contains all the observed vibrationally excited H$_2$, but the total column density of excited H$_2$, $N$(H$_2$*) $\la$ 10$^{19}$ cm$^{-2}$, does not appreciably increase the local $f$(H$_2$).}
 
\item{The column densities of CH$^+$ and CH are both very high (relative to the foreground) in the local material, however, and the ratio $N$(CH)/$N$(H$_2$) is much higher than the average value found for diffuse molecular gas in the Milky Way, where CH and H$_2$ are typically well correlated (Danks et al. 1984; Welty et al. 2006; Sheffer et al. 2008).
The small $f$(H$_2$), the high $N$(CH)/$N$(H$_2$) ratio, the strong CH$^+$, and the non-detection of CN in the local material near Herschel 36 suggest that most of the CH and CH$^+$ there may be produced non-thermally (Zsarg\'{o} \& Federman 2003).
While shocks may be responsible for at least some of the excited H$_2$ emission seen in the vicinity (Burton 2002), they probably are not responsible for the strong CH and CH$^+$ observed toward Herschel 36, as the velocities of the CH and CH$^+$ are very similar to the velocities of both the various atomic species toward Herschel 36 and the atomic and molecular lines seen toward the other neighboring stars in NGC 6530.}
 
\item{While the DIBs are stronger toward Herschel 36 than toward the other stars in NGC 6530, the ratios $W$(DIB)/$N$(H) for the material local to Herschel 36 are lower than the values typically found in the local Galactic ISM (Friedman et al. 2011), reminiscent of similar low values found in the Trapezium region (Herbig 1993; Friedman et al. 2011).  
The relatively low $f$(H$_2$), $N$(\ion{K}{1})/$N$(H$_{\rm tot}$)$^2$, and $W$(5780.5)/$N$(H) in the material local to Herschel 36 are suggestive of an elevated optical and UV radiation field, as expected from Herschel 36 and the other early-type stars in NGC 6530 (e.g., Herbig 1993; Welty \& Hobbs 2001; Vos et al. 2011).}
\end{itemize}

The three sight lines in Orion listed in Table~\ref{tab:local} (especially HD 37021 and HD 37061) exhibit some properties similar to those found for the local material near Herschel 36.
The UV extinction curves for the two Trapezium region stars are also fairly flat in the far-UV, with weak 2175 \AA\ bumps, and the values of $R_{\rm v}$ are high; the UV extinction and $R_{\rm v}$ for HD 37903 are closer to the Galactic averages, however (Fitzpatrick \& Massa 2007).
All three of the Orion sight lines have very low abundances of trace neutral species, but only the two Trapezium region sight lines also have low values of the $W$(5780.5)/$N$(H) ratio.
All three of the Orion sight lines exhibit absorption from vibrationally excited H$_2$, with the strongest absorption (apparently comparable to that toward Herschel 36) seen toward HD 37903 (Meyer et al. 2001; B. Rachford et al., in preparation).
In contrast to the local material near Herschel 36, the sight line to HD 37903 has fairly high $N$(H$_2$) and $f$(H$_2$), and a lower than average $N$(CH)/$N$(H$_2$) ratio.

The combination of high $N$(H) and relatively small size apparently characterizing the local region near Herschel 36 suggests that the mean density there must be moderately high.
If $N$(H) $\sim$ 7 $\times$ 10$^{21}$ cm$^{-2}$ (Table~\ref{tab:local}), and if an upper limit to the thickness of the region is adopted as 0.6 pc (the minimum distance to the nearest neighbor CD$-$24 13810), then the density $n$(H) would be $\sim$ 4000 cm$^{-3}$.  
If the thickness were as small as 400 AU (the minimum distance to Her 36 SE), however, then $n$(H) would be of order 10$^6$ cm$^{-3}$, close to the critical density of CH$^+$, and collisional effects would not be entirely negligible.
As such collisions would normally destroy the CH$^+$ even at much lower densities, a very rapid formation rate for CH$^+$ would be required (through the suggested non-thermal formation process), and perhaps formation of CH$^+$ in the excited state can occur (as H$_2$ is formed; see, e.g., Godard \& Cernicharo 2013).  
Since the excitation temperatures of CH, CH$^+$ and the $\lambda$5780 DIB differ toward Herschel 36 (Paper II), however, we cannot assume that they co-exist in a constant density region.  
It may be that we are not looking at material that can be represented by an average over distances of tenths of pc, but rather a region with condensations of material -- as are known in \ion{H}{2} regions and planetary nebulae from the proplyds and from the differences in abundances derived from permitted and forbidden lines (e.g., Liu 2011).  

The moderately high estimated mean densities and the indications of a significantly enhanced radiation field suggest that the local region may bear some resemblance to photodissociation regions (e.g., Hollenbach \& Tielens 1997, 1999).
Motivated by the recent observations of CH$^+$ emission obtained with {\it ISO} and {\it Herschel}, Godard \& Cernicharo (2013) have constructed models for the excitation of CH$^+$ in such regions, for ranges in density, temperature, and the strength of the ambient IR and optical radiation fields.
While the models appear able to reproduce the emission seen from the lowest few rotational levels of CH$^+$ toward NGC 7027 and the Orion Bar, the predicted $N$($J$=1)/$N$($J$=0) are typically of order 0.005 -- much lower than the value $\sim$0.2 seen toward Herschel 36 (Paper II).

\subsubsection{DIB Profiles from the Local Region near Herschel 36}
\label{sec-locdibs}

It would be important as well to disentangle the foreground DIB absorption from the DIB absorption arising from the local region near Herschel 36.  
As for the atomic and molecular species discussed above, we consider the DIBs observed toward the four nearest neighbors within NGC 6530 for which we have spectra.
The DIB equivalent widths and profiles seen toward those four stars are fairly similar (with some scatter in strength); the DIB profiles resemble those typically seen for other Galactic sight lines (with no extended redward wings).
As before, we assume that the profiles seen in the higher S/N spectra of 9 Sgr provide reasonable approximations to the foreground profiles, which can then be removed from the profiles observed toward Herschel 36, via subtraction of the corresponding optical depths at each wavelength, to estimate the DIB profiles for the small region near Herschel 36 where the DIB wings and the excited CH and CH$^+$ are produced.  
Because there could be small-scale fluctuations in the strengths of the foreground DIBs (in the R-C cloud and/or in material associated with NGC 6530), however, we cannot exclude the possibility that the DIBs seen toward 9 Sgr are not representative of the foreground material toward Herschel 36.
We therefore have subtracted foreground profiles spanning the range of DIB strengths seen in the various neighboring NGC 6530 and foreground sight lines listed in Table~\ref{tab:region}.

\begin{figure}[b!]
\epsscale{0.8}
\plotone{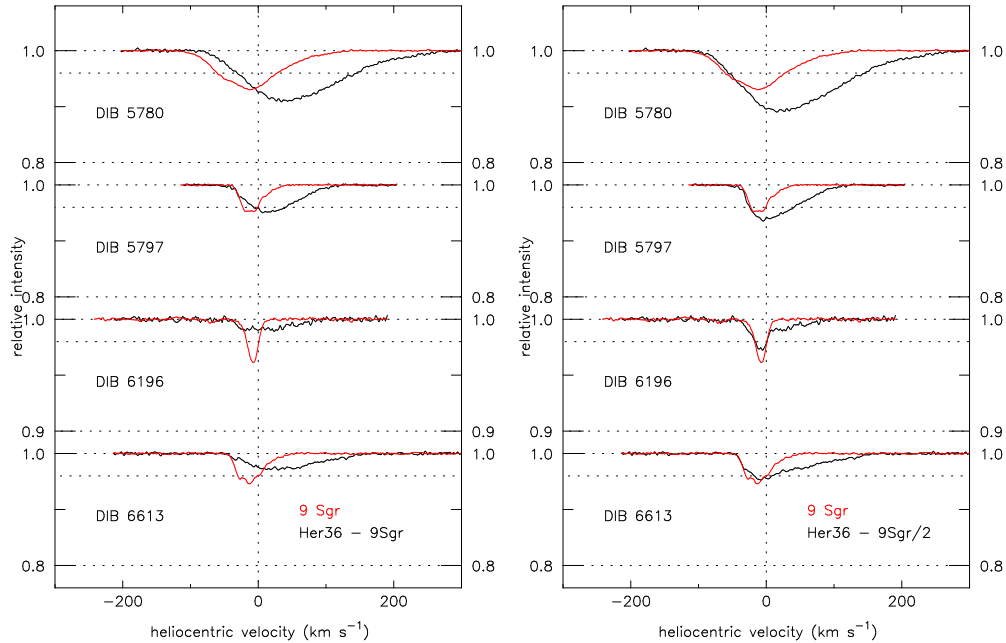}
\caption{Profiles of four DIBs:  as observed toward 9 Sgr (red) and two estimates for the local (foreground-subtracted) profiles toward Herschel 36 (black). 
On the left, the foreground DIBs toward Herschel 36 are assumed to be as seen toward 9 Sgr (3 arcmin away); on the right, the foreground DIBs are half as strong (likely a minimum for the foreground DIB strengths).
In each case, the subtraction is done for the (apparent) optical depths at each point.
For the profiles on the left, the sharp edge on the left disappears and the band center shifts to the right (longward) by 0.3 to 1.4 \AA, relative to 9 Sgr; the subtracted profiles generally are somewhat smoother and more bowl-shaped than the profiles observed toward Herschel 36.
For the profiles on the right, more of the core absorption remains -- more similar to the model profiles discussed in Paper II.}
\label{fig:diff}
\end{figure}

Figure~\ref{fig:diff} shows the foreground-subtracted profiles of our four primary DIBs for two choices of foreground DIB strength:  1) as seen toward 9 Sgr (representative of the average of the near neighbor sight lines within 6 arcmin of Herschel 36), and 2) half that strength.
The profiles on the left, where the full 9 Sgr profiles were subtracted, generally appear to be smoother, shallower, and shifted to the red by 0.4--1.4 \AA (relative to 9 Sgr), with significantly reduced absorption at the wavelengths of the normal DIB cores.
Similar results are obtained for most of the other DIBs exhibiting extended redward wings toward Herschel 36; only the $\lambda$6234.0 DIB retains a distinct core in the foreground-subtracted profile.
As discussed in Paper II, the ``coreless'' subtracted DIB profiles are difficult to understand.
If the foreground DIBs are actually weaker toward Herschel 36 than toward 9 Sgr, however, then more of the DIB cores would remain in the foreground-subtracted profiles -- as seen for the corresponding profiles on the right.
Given the ranges in $E(B-V)$, $N$(H), and DIB strengths seen for the neighboring NGC 6530 and foreground stars within 2 degrees of Herschel 36 (Table~\ref{tab:region}), the foreground DIBs are unlikely to be more than that factor of 2 weaker toward Herschel 36.
One could also speculate that emission from the DIB carriers in the surrounding nebulosity might ``fill in'' the DIB cores toward Herschel 36, but we have not detected the $\lambda$5797.1 DIB in emission from the Hourglass (\S~\ref{sec-mol}) -- and it is not clear that any such emission would be confined to the cores of the profiles.
For some of the other, weaker DIBs (with $W_\lambda$ $\la$ 5 m\AA), the equivalent widths are very similar toward Herschel 36 and 9 Sgr -- suggesting that those DIBs arise primarily in the foreground material, and are either very weak or absent in the local material near Herschel 36.
While no broad, shallow redward wings are seen for those weaker DIBs, such wings would be difficult to detect in the currently available spectra.
 
\subsubsection{Additional Observations Needed}
\label{sec-newobs}

Additional observations are needed to better understand both the physical conditions in the local region near Herschel 36 and how the DIB profiles behave under those conditions.  
Accurate determination of the width of the CH$^+$ lines would provide better constraints on the temperature and turbulence in the region of CH$^+$ formation.  
If the temperature of the local material is $\sim$ 1000 K, for example, the thermal width of the CH$^+$ line would be about $b_{\rm th}$ $\sim$ 1.1 km s$^{-1}$, which is slightly smaller than the $b$-value found to yield consistent $N$(CH$^+$) in fits to the CH$^+$ lines toward Herschel 36 (Paper II).  
A clear separation of local and foreground material and a clear temperature measurement would thus likely require a resolving power of order 0.5--1.0 km s$^{-1}$.  
Observations of the UV lines of several additional species should enable much better estimates for the density in the local region.
Measurement of the ground and excited fine-structure lines of \ion{C}{1} (below 1666 \AA, requiring {\it HST}/STIS) would yield the thermal pressure, and thus the hydrogen density (if the temperature is known; e.g., Jenkins \& Shaya 1979; Jenkins \& Tripp 2001).  
Also in that spectral region are many lines of vibrationally excited H$_2$, which could be used to determine level populations (and thus densities; e.g., Meyer et al. 2001; Gnaci\'{n}ski 2011) and line widths. 
At $T$ = 1000 K, for example, $b_{\rm th}$(H$_2$) $\sim$ 2.9 km s$^{-1}$, comparable to the 2.75 km s$^{-1}$ resolution (FWHM) of the STIS E140H grating.  
Accurate determination of the widths and velocities of the excited H$_2$ lines would help distinguish between shock and fluorescent excitation of the H$_2$ (Meyer et al. 2001; Burton 2002).
Observations of (intrinsically) strong UV absorption lines due to various neutral and ionized atomic species would provide tighter constraints on any possible components at positive velocities (i.e., overlapping the redward DIB wings).
The combination of very high resolution optical spectra near 3900--4300 \AA\ and a far-UV observation with {\it HST}/STIS would thus allow much better characterization of the physical conditions in the material local to Herschel 36 -- and thus of the cause of the extended redward DIB wings.
Finally, given the uncertainties regarding the foreground contributions to the DIB profiles toward Herschel 36, identification and detailed investigation of other sight lines showing similarly enhanced redward DIB wings -- but with less foreground contamination -- would enable tighter constraints to be placed on the behavior and identities of the DIB carriers.
  
\subsection{DIB Profile Variations in Other Sight Lines} 
\label{sec-prev}

There is a rich literature on DIB profiles and their variations in different lines of sight.  
A number of the DIBs (including those at 5780.5, 5797.1, and 6283.8 \AA) seen toward stars in the Orion Trapezium region are both broader and shifted slightly to the red (Porceddu et al. 1992; Kre{\l}owski \& Greenberg 1999).  
Examination of our APO/ARCES spectra of the DIBs in the Orion Trapezium region (toward HD 37021, HD 37022, HD 37061, and also HD 38087) indicates that (1) many of the DIBs are weak, relative to $N$(H) and/or $E(B-V)$, and (2) a number of the DIBs (e.g., those at 5705.1, 5780.5, 5797.1, 6204.8, and 6613.6 \AA) exhibit redshifts and/or extended redward wings.  
The $\lambda$5780.5 DIB, for example, appears to be both redshifted by 20--30 km s$^{-1}$ and broadened (in FWHM) by 20--30 km s$^{-1}$.  
Kre{\l}owski \& Greenberg (1999) ascribed those profile differences to accretion of the carriers onto grains, but the predicted weak, blueward emission wings were not seen.

Toward the runaway star HD 34078 (AE Aur), presently interacting with a diffuse molecular cloud (Boiss\'{e} et al. 2005, 2009), some of the narrower DIBs (e.g., those at 5797.1 and 6613.6 \AA) are blueshifted by $\sim$ 0.1 \AA\ ($\sim$ 5 km~s$^{-1}$; Galazutdinov et al. 2006).  
Both DIBs are broader than usual, with more prominent red wings than usual.  
The broader DIBs (e.g., those at 5780.5 and 6204.8 \AA) do not appear to be shifted, however.  
In our APO spectra, in addition to the slight (5--10 km s$^{-1}$) blueshifts of some of the narrower DIBs, some of the DIBs (e.g., those at 5705.1, 5797.1, 6196.0, 6204.8, 6379.2 and 6613.6 \AA) exhibit slightly extended redward wings.  
There may be a relation between this observation and the Herschel 36 phenomenon.  
Galazutdinov et al. (2008) found some DIBs longward of 5700 \AA\ to be blueshifted (by $\sim$ 6 km s$^{-1}$ for the $\lambda$6196.0 DIB) toward several stars in the Sco OB1 association -- but those shifts may just reflect the complex component structure (as seen in \ion{K}{1} and \ion{Na}{1} absorption) in those sight lines.  
Some of these DIBs also appear to be somewhat broader (generally due to redward wings) in several other sight lines in our data base that were not previously noted for profile variations (e.g., HD 281159, NGC 2264 Walker 67). 

Le Coupanec et al. (1999) proposed that the broadening of the $\lambda$5797.1 DIB in some reflection nebulae might be due to a higher rotational temperature of the carrier, but were unable to establish any correlation between the broadening and the gas--star distance; no broadening was seen for the $\lambda$6379.3 or $\lambda$6613.6 DIBs.  
These regions may have IR bright sources and effects similar to those noted here for Herschel 36 may be indicated.

Very high resolution spectra obtained by Galazutdinov et al. (2002) showed the narrow core of the $\lambda$6196.0 DIB to be up to 50 percent broader toward some stars in the Sco-Oph region, with evident substructure in the profiles.  
Galazutdinov et al. did not comment on a shallow wing to the red for that DIB, nor did they note any such effects for the broader $\lambda$5780.5 DIB.  
In our APO spectra, some of the DIBs (e.g., those at 5780.5, 5797.1, 6196.0, and 6613.6 \AA) in the more reddened sight lines in the rho Oph region (HD 147888, HD 147889, HD 147933) also have extended redward wings; the $\lambda$5780.5 DIB toward HD 147889, for example, is broadened by 15--20 km s$^{-1}$ (and slightly redshifted); see Figure~\ref{fig:meandib}.   
Ka\'{z}mierczak et al. (2009, 2010) have proposed a general correlation between the FWHM of the $\lambda$6196.0 DIB and the rotational excitation temperature of C$_2$, but that correlation does not appear to hold for a larger, more diverse sample of sight lines (D. Welty et al., in preparation). 

Examination of the ARCES spectra thus confirms many of the subtler profile variations previously discussed in the literature -- e.g., for sight lines in the Trapezium region, in Sco-Oph, and toward several other individual stars -- which may be related to the differences reported in this paper for Herschel 36.
In the cases for which wings have been noted, there are often indications of enhanced ultraviolet radiation fields (e.g., from low abundances of H$_2$ and/or trace neutral species), higher than average $R_{\rm v}$, shallow far-UV extinction, and/or abundant dust.  
Nebular emission lines often are apparent in the spectra.
In almost none of those other cases, however, are the variations as extreme as those seen toward Herschel 36 -- where (for example) the $\lambda$5780.5 DIB is redshifted by $\sim$ 45 km s$^{-1}$ and broadened by $\sim$ 60 km s$^{-1}$ (in FWHM), relative to the ``normal'' profile seen toward 9 Sgr.  
One possible exception is the DIB at 6204.8 \AA\ toward HD 37061, where the redward wing is more prominent than the one seen toward Herschel 36 (\S~\ref{sec-otherdibs}). 
 
\section{CONCLUSIONS}
\label{sec-conc}

The line of sight to the star Herschel 36, in the \ion{H}{2} region M8 near the Hourglass Nebula, exhibits several very unusual interstellar properties:  absorption lines from rotationally excited CH$^+$ and CH and from vibrationally excited H$_2$; an atypical extinction curve (with $R_{\rm v}$ $\sim$ 6, a weak 2175 \AA\ bump, and flat far-UV); and, for at least four of the stronger diffuse interstellar bands, remarkable extended redward wings that are much stronger and broader than have been reported toward any other star (to our knowledge). 
Because those unusual features are not present toward any of the neighboring stars in the cluster NGC 6530 that have been observed (with projected separations as small as 0.6 pc), they probably arise in a fairly small region, close to Herschel 36, which contains the bulk of the interstellar gas and dust in the sight line.
The material in that local region has a low fraction of hydrogen in molecular form [$f$(H$_2$) $\la$ 0.02], relatively low abundances of trace atomic species, and relatively low ratios of the equivalent widths of some DIBs to $N$(H) -- all suggestive of a fairly strong ambient UV radiation field.
Both CH and CH$^+$ exhibit strong absorption, however, and it is likely that both of those molecular species are formed non-thermally in the local gas.
For the DIBs exhibiting the extended redward wings, the profiles arising just from the material in the local region appear to exhibit weaker absorption at the wavelengths of the commonly observed DIB cores (together with the extended wings) -- but the strength of the core absorption depends on a somewhat uncertain correction for the contributions from unrelated foreground material.

The unique interstellar features seen toward Herschel 36 provide both new information on the behavior of the DIBs and new constraints on possible DIB carrier molecules.
As discussed in Paper II, a strong infrared point source, probably located within about 400 AU of Herschel 36, appears to be responsible for the excitation of CH and CH$^+$.
The broad redward DIB wings are then interpreted in terms of IR pumping of the closely spaced high-$J$ levels of relatively small (3--6 heavy atoms) DIB carrier molecules with strong permanent dipole moments.  
The requirement that the carriers of those DIBs must be polar molecules eliminates a number of suggested or possible candidate molecules from consideration.
The broad wings on those DIBs extend the normal redward intersection with the stellar continuum to 100--250 km s$^{-1}$ and more than double the region of the spectrum over which the continuum must be drawn to obtain accurate equivalent widths.  

Weaker, less extensive redward wings may be discerned (and some have been previously noted) for one or more of the DIBs in a few other sight lines (e.g., in the Orion Trapezium and Sco-Oph regions).
Such subtle profile variations may be common and may affect the measurements (and uncertainties) of the equivalent widths, and hence the precision of DIB--DIB correlation coefficients (McCall et al. 2010). 
For many of those sight lines, there are indications of enhanced ultraviolet radiation fields (e.g., from low abundances of H$_2$ and/or trace neutral species), higher than average $R_{\rm v}$, shallow far-UV extinction, and/or abundant dust -- i.e., similar to the case of Herschel 36.
Nebular emission lines often are apparent in the spectra.

There is no evidence that the unusual conditions in the local material near Herschel 36 lead to the creation of new DIBs that have not already been recognized in existing optical surveys, to limits of $\sim$ 10 m\AA\ for relatively narrow DIBs.
There is evidence, however, that some DIBs found in our DIB atlases for HD 204827 and HD 183143 are either much weaker or missing toward Herschel 36, but we have not discerned a pattern.  
Further work may indicate a new family of DIBs whose carriers are modified or destroyed in a strong UV field.

If the interpretation of the broadening as due to excited rotational levels in relatively simple molecules is correct, then the Cosmic Microwave Background Radiation would set the lower limit to the width of the DIBs, thus explaining the difference in width between atomic and diatomic interstellar lines and the ``diffuse'' interstellar bands.

\acknowledgments

We thank Alan Hirshfeld and Owen Gingerich for pointing out Herschel's 1847 volume as the source of the star's identification as Herschel 36, Steve Federman and Ed Jenkins for helpful discussions on the density of the CH$^+$ region, and Ben McCall for construction of the database which houses the ARCES DIB spectra (as well as much ancillary and derived information).  
Vivian Hoette was instrumental in obtaining the PROMPT images shown in Fig.~\ref{fig:prompt}; we also acknowledge the builders of the Skynet Robotic Telescope Network (D. Reichart, K. Ivarsen, J. Haislip) and of the PROMPT2 and PROMPT3 telescopes used for those observations (M. Nysewander, A. LaCluyze, K. Ivarsen).
Undergraduate students Zachary Taylor and Brianna Faint assisted JD in examining the ARCES spectra of Herschel 36.
We have made extensive use of the ESO public spectroscopic archive, which includes data from UVES, FEROS, and HARPS.  
Support for this work has been provided by the National Science Foundation, under grants AST-1009603 (DGY), AST-1009929 (TPS), AST-1008424 (JD), AST-1008801 (BLR), and AST-1238926 (DEW).

{\it Facilities:} \facility{ARC (ARCES)}, \facility{CTIO: PROMPT}, \facility{Magellan: Clay (MIKE)}, \facility{Max Planck: 2.2m (FEROS)}, \facility{VLT: Kueyen (UVES)}

\end{document}